\documentclass[twocolumn,showpacs,prb,superscriptaddress,a4paper,floatfix,showkeys]{revtex4}
\usepackage{graphicx}
\usepackage{amsmath}
\usepackage{graphicx}
\usepackage{dcolumn}
\usepackage{bm}

\begin{document}

\title{Decoherence and Thermalization of Quantum Spin Systems}
\author{Shengjun Yuan}
\affiliation{Institute of Molecules and Materials, Radboud University of Nijmegen, 6525
ED Nijmegen, The Netherlands}

\date{\today }

\begin{abstract}
In this review, we discuss the decoherence and thermalization of a quantum
spin system interacting with a spin bath environment, by numerically solving
the time-dependent Schr\"{o}dinger equation of the whole system. The effects
of the topologic structure and the initial state of the environment on the
decoherence of the two-spin and many-spin system are discussed. The role of
different spin-spin coupling is considered. We show under which conditions
the environment drives the reduced density matrix of the system to a fully
decoherent state, and how the diagonal elements of the reduced density
matrix approach those expected for the system in the microcanonical or
canonical ensemble, depending on the character of the additional integrals
of motion. Our demonstration does not rely on time-averaging of observables
nor does it assume that the coupling between system and bath is weak. Our
findings show that the microcanonical distribution (in each eigenenergy
subspace) and canonical ensemble are states that may result from pure
quantum dynamics, suggesting that quantum mechanics may be regarded as the
foundation of quantum statistical mechanics. Furthermore, our numerical
results show that a fully decoherent quantum system prefers to stay in an
equilibrium state with a maximum entropy, indicating the validity of the
second law of thermodynamics in the decoherence process.
\end{abstract}

\maketitle
\tableofcontents
\email{s.yuan@science.ru.nl}

\section{Introduction}

The manner in which a quantum system becomes effectively classical is of
great importance for the foundations of quantum physics. It has become
increasingly clear that the symptoms of classicality of quantum systems can
be induced by their environments~\cite{jens85}.

Intuitively, we expect that by turning on the interaction between the
quantum system and the environment, the fluctuations in the environment will
lead to a reduction of the coherence in the quantum system. This process is
called decoherence~\cite%
{Zeh1970,Zeh1996,Zurek1981,Zurek1982,Zurek1991,Zurek1998,Zurek2002,Zurek2003,Zurek2009,paz}%
. The existence of decoherence in the quantum system represents a challenge
for the realization of quantum computation and quantum information
processing~\cite{Nielsen2000,Kane1998,Loss2007,Leuenberger2001}, which are
expected to rely heavily on quantum coherence. In general, there are two
different mechanisms that contribute to decoherence. If the environment is
dissipative (or coupled to a dissipative system), the total energy is not
conserved and the whole system relaxes to a stationary equilibrium state,
for instance the thermal equilibrium state. In this review, we exclude this
class of dissipative processes and restrict ourselves to closed systems in
which a small quantum system is brought in contact with a much larger
environment. Then, decoherence is solely due to the fact that the initial
product state (wave function of the quantum system times wave function of
the environment) evolves into an entangled state of the whole system. The
interaction with the environment causes the initial pure state of the
quantum system to evolve into a mixed state, described by a reduced density
matrix \cite{neumann}, obtained by tracing out all the degrees of freedom of
the environment~\cite%
{Zeh1970,Zeh1996,Zurek1981,Zurek1982,Zurek1991,Zurek1998,Zurek2002,Zurek2003,Zurek2009,paz,feynman,leggett}%
.

The decoherence programme is supposed to explain the macroscopic quantum
superposition (\textquotedblleft Schr\"{o}dinger cat\textquotedblright )
paradox, that is, the inapplicability of the superposition principle to the
macroworld. The states that are \textquotedblleft robust\textquotedblright\
with respect to the interaction with the environment are called pointer
states~\cite{Zurek2003}. If the Hamiltonian of the quantum system $H_{S}$ is
a perturbation, relative to the interaction Hamiltonian $H_{int}$, the
pointer states are eigenstates of $H_{int}$~\cite{Zurek2003,Zurek2009,paz}.
In this case, the pointer states are essentially \textquotedblleft classical
states\textquotedblright , such as states with definite particle positions
or with definite spin directions of individual particles for magnetic
systems. In general, these classical \textquotedblleft Schr\"{o}dinger cat
states\textquotedblright , being a product state of individual particles
forming the system, are not entangled.

On the contrary, if the interaction Hamiltonian $H_{int}$ is a perturbation,
relative to the Hamiltonian of the quantum system $H_{S}$, the pointer
states are eigenstates of $H_{S}$ \cite{Zurek2003,paz}. In this case, the
pointer states are not necessary classical-type states, they may be
\textquotedblleft quantum\textquotedblright\ states such as standing waves,
stationary electron states in atoms, tunneling-split states for a particle
distributed between several potential wells, singlet or triplet states for
magnetic systems, etc.~\cite{paz}. This may explain, for example, that one
can observe linear atomic spectra - the initial states of an atom under the
equilibrium conditions are eigenstates of its Hamiltonian and not arbitrary
superposition thereof. Some less trivial pointer states have been found in
computer simulations of quantum spin systems for some range of the model
parameters~\cite{ourPRL,ourPRE}. In fact, the evolution to equilibrium of
quantum spin systems is still an open issue. Recently, the effect of an
environment of $N\gg 1$ spins on the entanglement of quantum spin systems
has attracted much attention~\cite%
{ourPRL,ourPRE,Yuan2006,Yuan2007,Yuan2008,Yuan2009,Melikidze2004,Wezel2005,Gao2005,YuanXZ2007,YuanXZ2005,ZhangW2006,ZhangW2007,Dawson2005,Rossini2007,Tessieri2003,Camalet2007,Bhaktavatsala2007,Relano2007,Dobrovitski2000,SKRO2006,ZhangGF2005,Hamdouni2006}%
.

Furthermore, recent simulation results of quantum spin systems show that
different statistical ensembles such as the microcanonical (per eigenenergy
subspace) and the canonical ensemble, could arise from the distribution of
these pointer states (eigenstates) in the mixed state\cite{Yuan2009}. This
direct connection of quantum dynamics and statistical mechanics gives an
explanation of a basic postulate in statistical mechanics:~a generic
\textquotedblleft system\textquotedblright\ that interacts with a generic
environment evolves into a state described by the canonical ensemble.
Experience shows that this is true but a detailed understanding of this
process, which is crucial for a rigorous justification of statistical
physics and thermodynamics, is still lacking\cite%
{balescu,Popescu2006,Rigol2008,Goldstein2006,Reimann2007,Reimann2008,Gemmer2006,Gemmer2006b, Cazalilla2006,Rigol2006, Rigol2007,Eckstein2008,Cramer2008,Cramer2008b,Flesch2008,Bocchieri1959,Shankar1985,Tasaki1998,sait96,Esposito2003,Merkli2007,Benenti2001}%
. In particular, in this context the meaning of \textquotedblleft
generic\textquotedblright\ is not clear. The key question is to what extent
the evolution to the equilibrium state depends on the details of the
dynamics of the whole system.

Earlier demonstrations of the fact that the system can be in the canonical
ensemble state are based on the Ergodic averages, that the expectation
values of the dynamical variables of the system approach their values for
the subsystem that is in the thermal equilibrium state~\cite%
{Bocchieri1959,Shankar1985,Tasaki1998,sait96}, or do not consider the
dynamics of the system but assume that the state of the whole system has a
special property called \textquotedblleft canonical
typicality\textquotedblright ~\cite%
{Popescu2006,Rigol2008,Goldstein2006,Reimann2007,Reimann2008,Gemmer2006,Gemmer2006b}%
. There are two basic assumptions in the derivation of the canonical
typicality, one is that the whole system is in the microcanonical ensemble,
another is that the interaction between the system and the environment is so
small that it can be neglected. The theory of the canonical typicality is
kinetic rather than dynamic, and it is yet unclear under which conditions
the whole system will evolve into the region in Hilbert space where its
subsystems are in the thermal equilibrium state \cite{Popescu2006}. A very
different setting to study nonequilibrium quantum dynamics is to start from
an eigenstate of some initial Hamiltonian and push the system out of this
state by a sudden change of the model parameters~\cite%
{Cazalilla2006,Rigol2006,Rigol2007,Eckstein2008,Cramer2008,Cramer2008b,Flesch2008}%
. To the best of our knowledge, it has not yet been shown that this approach
leads to the establishment of the canonical equilibrium distribution. We
also want to draw attention to the fact that a demonstration of relaxation
to the canonical distribution requires a system with at least three
different eigenenergies because a diagonal density matrix of a two-level
system can always be represented as a canonical distribution \cite%
{Esposito2003,Merkli2007}.

In this review, we focus on recent results concerning the decoherence and
thermalization in quantum spin systems \cite%
{ourPRL,ourPRE,Yuan2006,Yuan2007,Yuan2008,Yuan2009}. In section II, we give
a general theory about decoherence and thermalization, and introduce several
quantities that measure the effect of decoherence and thermalization. In
section III, we introduce the quantum spin model and the methods used in the
numerical simulation. In section IV, we focus on the decoherence of a
two-spin quantum system, and consider the effect of different topological
structures and different types of spin-spin coupling. In section V, we
discuss how a Heisenberg two-spin system evolves to the ground state. In
section VI, we consider the decoherence of many-spin systems. We show under
which conditions the environment drives the reduced density matrix of the
system to a fully decoherent state, which is described by the microcanonical
distribution per eigenenergy subspace. In section VII, by introducing the
energy dissipation of the many-spin system, we show that the diagonal
elements of the reduced density matrix approach those expected for the
system in the canonical ensemble. Section VIII contains the conclusion and a
brief discussion about the second law of thermodynamics in quantum systems.

\section{General Theory}

In general, the state of a closed quantum system is described by a density
matrix~\cite{Neumann55,BALL03} $\rho $. The state of a quantum system
interacting with an environment is represented by the reduced density matrix 
$\rho (t)$, obtained by tracing out all the degrees of freedom of the
environment. Decoherence of the quantum system means that the amplitude of
the off-diagonal terms in the reduced density matrix become smaller, and
full decoherence corresponds to all off-diagonal terms being zero. Here we
consider the case that the interaction Hamiltonian $H_{int}$ is a
perturbation, but not necessary very small, relative to the Hamiltonian of
the quantum system $H_{S}$, and the reduced density matrix of the quantum
system is represented in its energy eigenstates. The decoherence of the
quantum system can also be monitored by the time dependence of the quadratic
entropy $S_{S}\left( t\right) =1-Tr\rho ^{2}\left( t\right) $ and the
Loschmidt echo \cite{Cucchietti2003} $L\left( t\right) =Tr\left( \rho \left(
t\right) \rho _{0}\left( t\right) \right) $, where $\rho _{0}\left( t\right) 
$ is the density matrix for $H_{int}=0$.

The microcanonical ensemble is a mixed state where all accessible
eigenstates have equal weight. The microcanonical distribution per
eigenenergy subspace is characterized by a density matrix that is diagonal
with respect to the eigenstates of the system Hamiltonian, and the diagonal
elements which belong to the degenerate energy eigenstates are equal. A
state with microcanonical distribution per eigenenergy subspace is a
microcanonical ensemble if it has only one accessible eigenenergy.

The canonical ensemble is a mixed state where the diagonal elements take the
form $\exp (-\beta E_{i})$, $\beta =1/k_{B}T$ is proportional to the inverse
of the temperature $T$ ($k_{B}$ is Boltzmann's constant) and the $E_{i}$'s
denote the eigenenergies.

The distribution of the state of a quantum system is the microcanonical or
canonical ensemble only if it is in a fully decoherent state.

The time evolution of a closed quantum system is governed by the
time-dependent Schr{\"{o}}dinger equation (TDSE)~\cite{Neumann55,BALL03}. If
the initial density matrix of an isolated quantum system is non-diagonal,
then, according to the TDSE, its density matrix remains nondiagonal and
never approaches the thermal equilibrium state with the canonical or
microcanonical distribution. Therefore, in order to thermalize the system $S$%
, it is necessary to have the system $S$ interact with an environment ($E$),
also called the heat bath. Thus, the Hamiltonian of the whole system ($S+E$)
takes the form $H=H_{S}+H_{E}+H_{SE}$, where $H_{S}$ and $H_{E}$ are the
system and environment Hamiltonian, respectively, and $H_{SE}$ describes the
interaction between the system and the environment.

The state of the system $S$ is described by the reduced density matrix 
\begin{equation}
\rho (t)\equiv \mathbf{Tr}_{E}\rho _{{\small S+E}}\left( t\right) ,
\label{eq1}
\end{equation}%
where $\rho _{{\small S+E}}\left( t\right) $ is the density matrix of the
whole system at time $t$ and $\mathbf{Tr}_{E}$ denotes the trace over the
degrees of freedom of the environment.

The coherence of the system is conveniently characterized by $\sigma (t)$,
which is a global measure for the size of the off-diagonal terms of the
reduced density matrix, defined by 
\begin{equation}
\sigma (t)=\sqrt{\sum_{i=1}^{n-1}\sum_{j=i+1}^{n}\left\vert \rho
_{ij}(t)\right\vert ^{2}}.
\end{equation}%
Here $n$ denotes the dimension of the Hilbert space of system $S$ and $\rho
_{ij}(t)$ is the matrix element $(i,j)$ of the reduced density matrix $\rho $
in the representation that diagonalizes $H_{S}$. If $\sigma (t)=0$ the
system is in a state of full decoherence.

The difference between the state $\rho \left( t\right) $ and a
microcanonical distribution per eigenenergy subspace can be characterized by 
$\sigma (t)$ (measure of decoherence) and $\gamma (t)$ (measure of the
difference between the diagonal terms corresponding to the degenerate
eigenstates), which is defined by%
\begin{equation}
\gamma (t)=\sqrt{\sum_{i=1}^{n-1}\sum_{j=i+1}^{n}\left\vert \rho
_{ii}(t)-\rho _{jj}(t)\right\vert ^{2}\delta \left( E_{i}-E_{j}\right) }.
\end{equation}%
The system is in the microcanonical distribution per energy subspace if and
only if $\sigma (t)=0$ and $\gamma (t)=0.$

The system $S$ is in its thermal equilibrium state only if the reduced
density matrix takes the form 
\begin{equation}
\widehat{\rho }\equiv \left. {e^{-\beta H_{S}}}\right/ {\mathbf{Tr}%
_{S}e^{-\beta H_{S}}},  \label{eq2}
\end{equation}%
where $\mathbf{Tr}_{S}$ denotes the trace over the degrees of freedom of the
system $S$. The difference between the state $\rho \left( t\right) $ and the
canonical distribution $\widehat{\rho }$ is represented by $\sigma (t)$ and $%
\delta (t)$, defined by 
\begin{equation}
\delta (t)=\sqrt{\sum_{i=1}^{n}\left( \rho _{ii}(t)-\left. {e^{-b(t)E_{i}}}%
\right/ {\sum_{i=1}^{n}e^{-b\left( t\right) E_{i}}}\right) ^{2}},
\end{equation}%
with 
\begin{equation}
b(t)=\frac{\sum_{i<j,E_{i}\neq E_{j}}[\ln \rho _{jj}(t)-\ln \rho _{ii}(t)]/({%
E_{j}-E_{i}})}{\sum_{i<j,E_{i}\neq E_{j}}1}.
\end{equation}%
As the system relaxes to its canonical distribution both $\sigma (t)$ and $%
\delta (t)$ vanish, $b(t)$ converging to $\beta $.

\section{Quantum Spin System and Numerical Method}

Most theoretical investigations of decoherence have been carried out for
oscillator models of the environment for which powerful path-integral
techniques can be used to treat the environment analytically~\cite%
{feynman,leggett}. On the other hand, it has been pointed out that a
magnetic environment, described by quantum spins, is essentially different
from the oscillator model in many aspects~\cite{stamp}. For the simplest
model of a single spin in an external magnetic field, some analytical
results are known~\cite{stamp}. For the generic case of two and more spins,
numerical simulation~\cite%
{ourPRL,ourPRE,Melikidze2004,Yuan2006,Yuan2007,Yuan2008,Yuan2009} is the
main source of theoretical information.

\begin{figure}[t]
\begin{center}
\includegraphics[width=8.25cm]{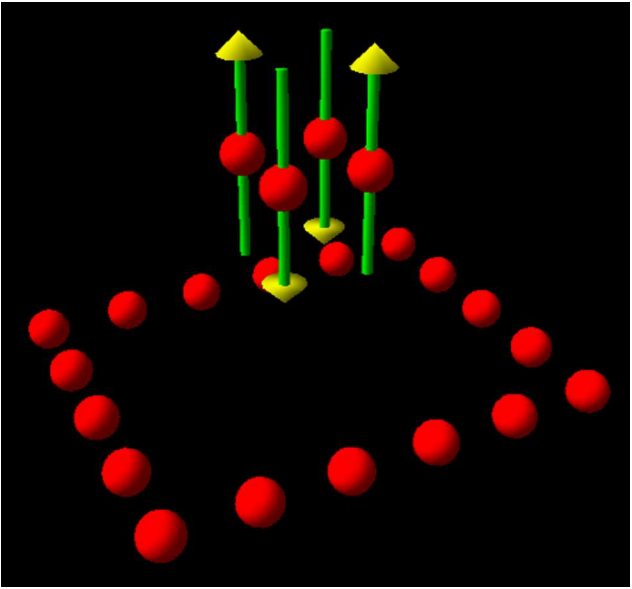}
\end{center}
\caption{A typical configuration of a quantum spin system surrounded by a
quantum spin bath ($N_{S}=4$ and $N=18$). The quantum spin system consists
of four spin $1/2$ particles, with orientation spin-up or spin-down. The
bath spins are in a complicated random superposition state for which the
expectation values $\left\langle S_{x}\right\rangle ,$ $\left\langle
S_{y}\right\rangle $ and $\left\langle S_{z}\right\rangle $ of each spin are
all zero.}
\label{spin3d}
\end{figure}

A generic quantum spin model can be described by the Hamiltonian $%
H=H_{S}+H_{E}+H_{SE}$ where%
\begin{eqnarray}
H_{S} &=&-\sum_{i=1}^{N_{S}-1}\sum_{j=i+1}^{N_{S}}\sum_{\alpha
=x.y,z}J_{i,j}^{\alpha }S_{i}^{\alpha }S_{j}^{\alpha },  \label{HAMS} \\
H_{E} &=&-\sum_{i=1}^{N-1}\sum_{j=i+1}^{N}\sum_{\alpha =x,y,z}\Omega
_{i,j}^{\alpha }I_{i}^{\alpha }I_{j}^{\alpha },  \label{HAME} \\
H_{SE} &=&-\sum_{i=1}^{N_{S}}\sum_{j=1}^{N}\sum_{\alpha =x,y,z}\Delta
_{i,j}^{\alpha }S_{i}^{\alpha }I_{j}^{\alpha }.  \label{HAMSE}
\end{eqnarray}%
Here the $S^{\alpha }$'s and $I^{\alpha }$'s denote the spin-1/2 operators
of the system and environment respectively (we use units such that $\hbar $
and $k_{B}$ are one). An analytic solution of the TDSE can only be obtained
for very special choices of the exchange integrals $J_{i,j}^{\alpha }$, $%
\Omega _{i,j}^{\alpha }$ and $\Delta _{i,j}^{\alpha }$ but it is
straightforward to solve the TDSE numerically for any choice of the model
parameters.

The state, that is the density matrix $\rho (t)$ of the whole system at time 
$t$, is completely determined by the choice of the initial state of the
whole system and the numerical solution of the TDSE. In this review, the
initial state of the whole system (S+E) is a pure state. This state evolves
in time according to 
\begin{equation*}
|\Psi (t)\rangle =e^{-iHt}|\Psi (0)\rangle
=\sum_{i=1}^{n_{S}}\sum_{p=1}^{n}c(i,p,t)|i,p\rangle ,
\end{equation*}%
where the states $\{|i,p\rangle \}$ denote a complete set of orthonormal
states. In terms of the expansion coefficients $c(i,p,t)$, the reduced
density matrix reads 
\begin{eqnarray}
\rho (t)_{i,j} &=&\mathbf{Tr}_{E}\sum_{p=1}^{n}\sum_{q=1}^{n}c^{\ast
}(i,q,t)c(j,p,t)|j,p\rangle \langle i,q|  \notag \\
&=&\sum_{p=1}^{n}c^{\ast }(i,p,t)c(j,p,t),  \label{eq5}
\end{eqnarray}%
which is easy to compute from the solution of the TDSE. We monitor the
effects of decoherence by computing the matrix elements of the reduced
density matrix $\rho (t)$ of the quantum system. As explained earlier, in
the regime of interest $|\Delta |\ll |J|$, the pointer states are expected
to be the eigenstates of the quantum systems. Hence, we compute the matrix
elements of the density matrix in the basis of eigenvectors of the quantum
system. We also compute the time dependence of the quadratic entropy $%
S_{S}\left( t\right) =1-Tr\rho ^{2}\left( t\right) $ and the Loschmidt echo 
\cite{Cucchietti2003}$L\left( t\right) =Tr\left( \rho _{0}\left( t\right)
\rho \left( t\right) \right) $, where $\rho _{0}\left( t\right) $ is the
density matrix for $H_{SE}=0.$ Another quantity of interest that can be
extracted from the solution of the TDSE is the local density of states
(LDOS) 
\begin{eqnarray}
\mbox{LDOS}(E) &\equiv &\frac{1}{2\pi }\int_{-\infty }^{+\infty
}dt\;e^{-iEt}\langle \Psi (0)|e^{-iHt}|\Psi (0)\rangle  \notag  \label{eq6}
\\
&=&\sum_{k=1}^{D}|\langle \Psi (0)|\varphi _{k}\rangle |^{2}\delta (E-E_{k}),
\end{eqnarray}%
where $D=n+n_{S}$, $\{|\varphi _{k}\rangle \}$, and $\{E_{k}\}$ denote the
dimension of the Hilbert space, the eigenstates and eigenvalues of the whole
system, respectively. The LDOS is \textquotedblleft local\textquotedblright\
with respect to the initial state: It provides information about the overlap
of the initial state and the eigenstates of $H$. 
%A detailed discussion of the computational of the LDOS can be found in Ref.~\cite{HAMS00}.

The notation to specify the initial state is as follows:

(1) $\left\vert GROUND\right\rangle _{S}$ is the ground state or a random
superposition of all degenerated ground states of the system;

(2) $\widetilde{\left\vert GROUND\right\rangle }_{S}$ is the state which has
the energy close but not equal to that of the ground state;

(3) $\left\vert UU\right\rangle _{S}$ is a state in which all spins of the
system are up, meaning that in this state, the expectation value of each
spin is one;

(4) $\left\vert UD\right\rangle _{S}$ is a state in which two
nearest-neighbor spins of the system are antiparallel implying that in this
state, the correlation of their $z$-components is minus one;

(5) $\widetilde{\left\vert UD\right\rangle }_{S}$ is a state close to $%
\left\vert UD\right\rangle _{S}$, but the correlation of their $z$%
-components is larger than minus one;

(6) $\left\vert RR\right\rangle _{S}$ denotes the product state of random
superpositions of the states of the individual spins of the system;

(7) $\left\vert RANDOM\right\rangle _{S}$ denotes a random superposition of
all possible basis states.

The same notation is used for the spins in the environment, the subscript $S$
being replaced by $E$.

As we report results for many different types of spin systems it is useful
to introduce a simple terminology to classify them according to symmetry and
connectivity. The terms \textquotedblleft XY\textquotedblright ,
\textquotedblleft Heisenberg\textquotedblright , \textquotedblleft
Heisenberg-type\textquotedblright , \textquotedblleft
Ising\textquotedblright , \textquotedblleft Ising-type\textquotedblright\
and Ising$\pm $ system refer to the following cases:

(1) XY: $J_{i,j}^{x}=J_{i,j}^{y}=J$ and $J_{i,j}^{z}=0$;

(2) Heisenberg: $J_{i,j}^{x}=J_{i,j}^{y}=J_{i,j}^{z}=J;$

(3) Heisenberg-type: $J_{i,j}$ are uniform random in the range $[-\left\vert
J\right\vert ,\left\vert J\right\vert ]$;

(4) Ising: $J_{i,j}^{x}=J_{i,j}^{y}=0$ and $J_{i,j}^{z}=J$;

(5) Ising-type: $J_{i,j}^{x}=J_{i,j}^{y}=0$ and $J_{i,j}^{z}$ are uniform
random in the range $[-\left\vert J\right\vert ,\left\vert J\right\vert ]$;

(6) Ising$\pm :$ $J_{i,j}^{x}=J_{i,j}^{y}=0$ and $J_{i,j}^{z}$ are random $%
-\left\vert J\right\vert $ or $\left\vert J\right\vert $.

The same terminology is used for the Hamiltonian $H_{E}$ of the environment
and for the interaction Hamiltonian $H_{SE}$. In our model, all the spins of
the system interact with each spin of the environment. To characterize the
connectivity $K$ of the spins within the system (environment), we use the
term \textquotedblleft ring\textquotedblright\ ($K=2$) for spins forming a
one-dimensional chain with nearest-neighbor interactions and periodic
boundary conditions, \textquotedblleft square-lattice\textquotedblright\ ($%
K=4$)\ or \textquotedblleft triangular-lattice\textquotedblright\ ($K=6$) if
the spins are located on a two-dimensional square or triangular lattice with
nearest-neighbor interactions, and \textquotedblleft spin
glass\textquotedblright\ ($K=N-1$) when all the spins within the system
(environment) interact with each other with a Heisenberg-type interaction~%
\cite{Binder1986,Mezard1987}.

The time evolution of the whole system is obtained by solving the TDSE for
the many-body wave function $|\Psi (t)\rangle $, describing the system plus
the environment~\cite%
{Dobrovitski2003,Trotter1959,Suzuki1977,Suzuki1985,Suzuki1991,DeRaedt1983,DeRaedt1987,DeRaedt1989,Vries1993,Michielsen1998,DeRaedt2000,DeRaedt2006,Krech1998}%
. The numerical method that we use is described in Ref.~\cite%
{Dobrovitski2003}. It is based on the numerically exact Chebyshev polynomial
decomposition of the operator $U\left( t\right) =e^{-itH}.$ It is very
efficient and conserves the energy of the whole system to machine precision.
It is widely used in the simulation of quantum spin systems, such as the
study of decoherence \cite{ourPRL,ourPRE,Yuan2006,Yuan2007,Yuan2008,Yuan2009}%
, the modeling of a quantum computer \cite{DeRaedt2006}, the propagation of
the quantum spin wave \cite{Yuan2006a}, and the study of stability of the
quantum domain wall \cite{Yuan2007a}, etc..

The simulation of the wave function is performed in the spin-up and
spin-down basis, and it is convenient to use this basis to apply the
Hamiltonian operator on the wave vector, which is a basic operation in the
Chebyshev polynomial algorithm. Based on the wave function of the whole
system in the up-down basis, we first calculate the reduced density matrix
of the quantum system by tracing out the degrees of freedom of the
environment, then diagonalize the Hamiltonian $H_{S}$ to get all the energy
eigenvalues and the corresponding eigenstates of the quantum system, and
finally transform the reduced density matrix from the up-down basis to the
energy eigenstate basis. With the reduced density matrix we can calculate
all the physical quantities of the quantum system, such as the energy, the
entropy, the measure of decoherence $\sigma $, the measure of the
distribution $\gamma $, $\delta $ and the effective inverse temperature $b$,
etc.

\section{Decoherence of a Two-Spin System}

\begin{table*}[t]
\caption{The values of the correlation functions $\langle \mathbf{S}%
_{1}\cdot \mathbf{S}_{2}\rangle $, $\langle S_{1}^{z}S_{2}^{z}\rangle $, $%
\langle S_{1}^{x}S_{2}^{x}\rangle $, the total magnetization $M$, the
concurrence $C$ and the magnetization $\langle S_{1}^{z}\rangle $ for
different states of the two-spin quantum system.}
\label{table1}
\begin{center}
\begin{ruledtabular}
\begin{tabular}{ccccccc}
% \begin{tabular}{|c|c|c|c|c|}
% \hline
$|\varphi \rangle $ & $\langle \mathbf{S}_{1}\cdot \mathbf{S}_{2}\rangle $ &
$\langle S_{1}^{z}S_{2}^{z}\rangle$ & $\langle S_{1}^{x}S_{2}^{x}\rangle$ & $M$ & $C$ &
$\langle S_{1}^{x}\rangle$\\ \hline
$\left( \left\vert \uparrow \downarrow \right\rangle
-\left\vert \downarrow \uparrow \right\rangle \right) /\sqrt{2}$ & $-3/4$ & $-1/4$ & $-1/4$ & $0$ & $1$ & $0$ \\
$\left( \left\vert \uparrow \downarrow \right\rangle
+\left\vert \downarrow \uparrow \right\rangle \right) /\sqrt{2}$ & $1/4$ & $-1/4$ & $1/4$ & $0$ & $1$ & $0$ \\
$\left( \left\vert \uparrow \uparrow \right\rangle
-\left\vert \downarrow \downarrow \right\rangle \right) /\sqrt{2}$ & $1/4$ & $1/4$ & $-1/4$ & $0$ & $1$ & $0$ \\
$\left( \left\vert \uparrow \uparrow \right\rangle
+\left\vert \downarrow \downarrow \right\rangle \right) /\sqrt{2}$ & $1/4$ & $1/4$ & $1/4$  & $0$ & $1$ & $0$ \\
$\left\vert \uparrow \downarrow \right\rangle$ & $-1/4$ & $-1/4$ & $0$ & $0$ & $0$ & $1/2$\\
$\left\vert \downarrow \uparrow \right\rangle$ & $-1/4$ & $-1/4$ & $0$ & $0$ & $0$ & $-1/2$\\
$\left\vert \uparrow \uparrow \right\rangle $& $1/4$ & $1/4$ & $0$ & $1$ & $0$ & $1/2$\\
$\left\vert \downarrow \downarrow \right\rangle$ & $1/4$ & $1/4$ & $0$ & $-1$ & $0$ & $-1/2$
\end{tabular}
\end{ruledtabular}
\end{center}
\end{table*}

\subsection{A Two-Spin System}

The quantum state of a two-spin system is completely determined by its
reduced $4\times 4$ density matrix. Although the reduced density matrix
contains all the information about the quantum system, it is often
convenient to characterize the state of the two-spin system by other
quantities such as the correlation functions $\langle \mathbf{S}_{1}\cdot 
\mathbf{S}_{2}\rangle $, $\langle S_{1}^{z}S_{2}^{z}\rangle $, and $\langle
S_{1}^{x}S_{2}^{x}\rangle $, the single-spin magnetizations $\langle
S_{1}^{x}\rangle $, $\langle S_{2}^{x}\rangle $, and $M\equiv \langle
S_{1}^{z}+S_{2}^{z}\rangle $, and the concurrence $C$~\cite%
{Wootters97,Wootters98}. In Table~\ref{table1}, we show the values of these
quantities for different states of the two-spin system.

The concurrence, which is a convenient measure for the entanglement of the
two spins is defined as~\cite{Wootters97,Wootters98} 
\begin{equation}
C\left( \rho \right) =\max (0,\lambda _{1}-\lambda _{2}-\lambda _{3}-\lambda
_{4}),
\end{equation}%
where the $\lambda _{i}$ are the eigenvalues, in decreasing order, of the
Hermitian matrix 
\begin{equation}
R\equiv \sqrt{\sqrt{\rho }\widetilde{\rho }\sqrt{\rho }}.
\end{equation}%
Here $\rho $ is the reduced density matrix of central spin pairs based on
the standard basis $\left\vert \uparrow \right\rangle \left\vert \uparrow
\right\rangle $, $\left\vert \uparrow \right\rangle \left\vert \downarrow
\right\rangle $, $\left\vert \downarrow \right\rangle \left\vert \uparrow
\right\rangle $, $\left\vert \downarrow \right\rangle \left\vert \downarrow
\right\rangle $, and 
\begin{equation}
\widetilde{\rho }=(\sigma _{y}\otimes \sigma _{y})\rho ^{\ast }(\sigma
_{y}\otimes \sigma _{y}),
\end{equation}%
where $\sigma _{y}=\left( 
\begin{array}{cc}
0 & -i \\ 
i & 0%
\end{array}%
\right) $ and $\rho ^{\ast }$ is the complex conjugate of $\rho $.

In fact, the concurrence $C$ is a measure between the state $\left\vert \psi
\right\rangle $ and the state with the two spins flipped $\left\vert 
\widetilde{\psi }\right\rangle $:%
\begin{equation}
C=\left\vert \left\langle \psi |\widetilde{\psi }\right\rangle \right\vert .
\end{equation}%
The singlet state, $\left\vert \psi \right\rangle =\left( \left\vert
\uparrow \right\rangle \left\vert \downarrow \right\rangle -\left\vert
\downarrow \right\rangle \left\vert \uparrow \right\rangle \right) /\sqrt{2}$
is unchanged under flipping two spins, therefore $C=1$. The triplet state $%
\left\vert \psi \right\rangle =\left( \left\vert \uparrow \right\rangle
\left\vert \downarrow \right\rangle +\left\vert \downarrow \right\rangle
\left\vert \uparrow \right\rangle \right) /\sqrt{2}$ is also unchanged under
flipping two spins, so $C=1$. For $\left\vert \uparrow \right\rangle
\left\vert \uparrow \right\rangle $, $\left\vert \uparrow \right\rangle
\left\vert \downarrow \right\rangle $, $\left\vert \downarrow \right\rangle
\left\vert \uparrow \right\rangle $, and $\left\vert \downarrow
\right\rangle \left\vert \downarrow \right\rangle $, the state is totally
different if the two spins flip and then $C=0.$

In the case that the two spins are coupled by the isotropic Heisenberg
interaction, the Hamiltonian of the system simplifies%
\begin{equation}
H_{S}=-J\mathbf{S}_{1}\mathbf{S}_{2},
\end{equation}%
and the four eigenstates of $H_{S}$ are given by 
\begin{eqnarray}
\left\vert T_{1}\right\rangle &=&\left\vert \uparrow \uparrow \right\rangle
=\left\vert 1\right\rangle ,  \notag  \label{eigenstates} \\
\left\vert S\right\rangle &=&\frac{\left\vert \uparrow \downarrow
\right\rangle -\left\vert \downarrow \uparrow \right\rangle }{\sqrt{2}}%
=\left\vert 2\right\rangle ,  \notag \\
\left\vert T_{0}\right\rangle &=&\frac{\left\vert \uparrow \downarrow
\right\rangle +\left\vert \downarrow \uparrow \right\rangle }{\sqrt{2}}%
=\left\vert 3\right\rangle ,  \notag \\
\left\vert T_{-1}\right\rangle &=&\left\vert \downarrow \downarrow
\right\rangle =\left\vert 4\right\rangle ,
\end{eqnarray}%
satisfying%
\begin{equation}
H_{S}\left\vert S\right\rangle =E_{S}\left\vert S\right\rangle ,\text{ }%
H_{S}\left\vert T_{1,0,-1}\right\rangle =E_{T}\left\vert
T_{1,0,-1}\right\rangle ,
\end{equation}%
where $E_{S}=3J/4$ and $E_{T}=-J/4$.

From Table~\ref{table1}, it is clear that the singlet state $\left\vert
S\right\rangle $ is the most easily distinguished state as the two-spin
system is in the singlet state if and only if $\langle \mathbf{S}_{1}\cdot 
\mathbf{S}_{2}\rangle =-3/4$. To identify the other states, we usually need
to know at least two of the quantities listed in Table \ref{table1}. For
example, to make sure that the system is in the triplet state $\left\vert
T_{0}\right\rangle $, the values of $\langle \mathbf{S}_{1}\cdot \mathbf{S}%
_{2}\rangle $ and $\langle S_{1}^{z}S_{2}^{z}\rangle $ should match with the
corresponding entries of Table~\ref{table1}. Likewise, the two-spin system
will be in the state $\left\vert \uparrow \uparrow \right\rangle $ if $%
\langle \mathbf{S}_{1}\cdot \mathbf{S}_{2}\rangle =1/4$ and $M=1$.

If the interaction between the Heisenberg quantum system and the environment
is isotropic, that is, if $\Delta _{i,j}^{(x)}=\Delta _{i,j}^{(y)}=\Delta
_{i,j}^{(z)}\equiv \Delta $ for all $i,j$, then the Hamiltonian $H_{SE}$ is
simplified as 
\begin{equation}
H_{SE}=-\Delta (\mathbf{S}_{1}+\mathbf{S}_{2})\cdot \sum_{j=1}^{N}\mathbf{I}%
_{j},  \label{Hce1}
\end{equation}%
which leads to $[H_{S},H_{SE}]=0$. As shown in Ref. \cite{Yuan2007}, if the
energy of the quantum system is conserved, then the decoherence process is
determined by $H_{SE}$, $H_{E}$, the initial state of whole system $%
\left\vert \Psi (t_{0})\right\rangle $, and the eigenstates of the quantum
system. In other words, in this case, $L\left( t\right) $ and $\left\vert
\rho \left( t\right) \right\vert $ do not depend on $J$, which means that
the relative value of $\Delta /J$ has no effect on the decoherence process.
Furthermore, if we take the interactions between the environment spins to be
isotropic, that is, $\Omega _{i,j}^{(x)}=\Omega _{i,j}^{(y)}=\Omega
_{i,j}^{(z)}\equiv \Omega _{i,j}$ for all $i,j$, then the Hamiltonian 
\begin{equation}
H_{E}=-\sum_{i=1}^{N-1}\sum_{j=i+1}^{N}\Omega _{i,j}\mathbf{I}_{i}\cdot 
\mathbf{I}_{j}  \label{He1}
\end{equation}%
commutes with $H_{SE}$, and therefore $H_{E}$ has also no effect on the
decoherence process.

In fact, since $[H_{S},H_{SE}]=0$, the time evolution operator of the whole
system $e^{-iHt}$ can be represented as $e^{-iH_{S}t}e^{-i\left(
H_{SE}+H_{E}\right) t}.$ The initial state of the quantum system can be
represented as $\left\vert \varphi (t_{0})\right\rangle
=\sum_{k}a_{k}\left\vert k\right\rangle $, where $\{\left\vert
k\right\rangle \}$ and $\{E_{k}\}$ are the eigenstates and corresponding
eigenvalues of the quantum system, that is, $H_{S}\left\vert k\right\rangle
=E_{k}\left\vert k\right\rangle $.

For an isolated quantum system ($H_{SE}=0$), the time evolution of the
density matrix of the quantum system is given by 
\begin{equation}
\rho _{0}\left( t\right) =\sum_{k,l}e^{-i\left( E_{k}-E_{l}\right)
t}a_{k}a_{l}^{\ast }\left\vert k\right\rangle \left\langle l\right\vert .
\end{equation}%
If the quantum system is coupled to a bath with initial state $\phi \left(
t_{0}\right) $, the state of the whole system at time $t$ is given by 
\begin{eqnarray}
\left\vert \Psi (t)\right\rangle &=&e^{-iHt}\left\vert \Psi
(t_{0})\right\rangle  \notag \\
&=&\sum_{k}e^{-iE_{k}t}a_{k}e^{-i\left( H_{SE}+H_{E}\right) t}\left\vert
k\right\rangle \left\vert \phi \left( t_{0}\right) \right\rangle .
\end{eqnarray}%
As $[H_{S},H_{SE}]=0$, we have $H_{SE}\left\vert k\right\rangle \left\vert
\phi \left( t_{0}\right) \right\rangle =\left\vert k\right\rangle
M_{k}\left\vert \phi \left( t_{0}\right) \right\rangle $ and hence, the
state at time $t$ becomes%
\begin{equation}
\left\vert \Psi (t)\right\rangle =\sum_{k}a_{k}e^{-iE_{k}t}\left\vert
k\right\rangle \left\vert \phi _{k}\left( t\right) \right\rangle ,
\end{equation}%
where 
\begin{equation}
\left\vert \phi _{k}\left( t\right) \right\rangle \equiv e^{-i\left(
M_{k}+H_{E}\right) t}\left\vert \phi \left( t_{0}\right) \right\rangle .
\label{phit}
\end{equation}%
The density matrix $\rho _{S+E}\left( t\right) $ of the whole system is%
\begin{eqnarray}
\rho _{S+E}\left( t\right) &=&\left\vert \Psi (t)\right\rangle \left\langle
\Psi (t)\right\vert  \notag \\
&=&\sum_{k,l}e^{-i\left( E_{k}-E_{l}\right) t}a_{k}a_{l}^{\ast }\left\vert
k\right\rangle \left\vert \phi _{k}\left( t\right) \right\rangle
\left\langle l\right\vert \left\langle \phi _{l}\left( t\right) \right\vert ,
\notag \\
&&
\end{eqnarray}%
and the reduced density matrix $\rho \left( t\right) $ of the quantum system
is%
\begin{eqnarray}
\rho \left( t\right) &=&Tr_{E}\rho _{S+E}\left( t\right)  \notag \\
&=&\sum_{k,l}e^{-i\left( E_{k}-E_{l}\right) t}a_{k}a_{l}^{\ast }\left\langle
\phi _{l}\left( t\right) |\phi _{k}\left( t\right) \right\rangle \left\vert
k\right\rangle \left\langle l\right\vert .  \notag \\
&&
\end{eqnarray}%
The Loschmidt echo $L\left( t\right) $ of the quantum system can be
calculated as 
\begin{eqnarray}
L\left( t\right) &=&Tr\left( \rho \left( t\right) \rho _{0}\left( t\right)
\right)  \notag \\
&=&Tr{[}\sum_{k,l}e^{-i\left( E_{k}-E_{l}\right) t}a_{k}a_{l}^{\ast
}\left\langle \phi _{l}\left( t\right) |\phi _{k}\left( t\right)
\right\rangle \left\vert k\right\rangle \left\langle l\right\vert  \notag \\
&&\times \sum_{m,n}e^{-i\left( E_{m}-E_{n}\right) t}a_{m}a_{n}^{\ast
}\left\vert m\right\rangle \left\langle n\right\vert {]}  \notag \\
&=&\sum_{k,l}\left\vert a_{k}\right\vert ^{2}\left\vert a_{l}\right\vert
^{2}\left\langle \phi _{l}\left( t\right) |\phi _{k}\left( t\right)
\right\rangle .
\end{eqnarray}

It is clear that if $[H_{S},H_{SE}]=0$, the decoherence process is
determined by the initial state of the quantum system $\{a_{k}\}$ and the
time evolution of $\{\left\vert \phi _{k}\left( t\right) \right\rangle \}$.
As shown in Eq.~(\ref{phit}), the $\{\left\vert \phi _{k}\left( t\right)
\right\rangle \}$ are determined by the initial state of the bath $\phi
\left( t_{0}\right) $, the eigenstates $\{\left\vert k\right\rangle \}$ of
the quantum system, and the Hamiltonians $H_{SE}$ and $H_{E}$. The
eigenvalues $\{E_{k}\}$ have no effect on the decoherence process. Thus,
multiplying $H_{S}$ by a constant does not change $L\left( t\right) $ and the
diagonal elements of the reduced density matrix $\rho _{S}\left( t\right) $.
The time evolution of the absolute value of the off-diagonal elements 
\begin{equation}
\left\vert \rho _{S}\left( t\right) _{kl}\right\vert =\left\vert
a_{k}a_{l}^{\ast }\right\vert \left\langle \phi _{l}\left( t\right) |\phi
_{k}\left( t\right) \right\rangle ,
\end{equation}%
is independent of $H_{S}$. This means that the relevant values of the
coupling between the system spins ($J$) and between the system and the
environment ($\Delta $) have no effect on the decoherence process.

If $[H_{S},H_{SE}]=0$ and $[H_{SE},H_{E}]=0$, then, Eq.~(\ref{phit}) becomes%
\begin{equation}
\left\vert \phi _{k}\left( t\right) \right\rangle
=e^{-iM_{k}t}e^{-iH_{E}t}\left\vert \phi \left( t_{0}\right) \right\rangle ,
\end{equation}%
and therefore we have%
\begin{equation}
\left\langle \phi _{l}\left( t\right) |\phi _{k}\left( t\right)
\right\rangle =\left\langle \phi \left( t_{0}\right) \right\vert e^{-i\left(
M_{k}-M_{l}\right) t}\left\vert \phi \left( t_{0}\right) \right\rangle ,
\end{equation}%
implying that $\left\vert \rho _{S}\left( t\right) _{kl}\right\vert $ and $%
L\left( t\right) $ do not depend on $H_{E}$.

Our goal is to find under which conditions the quantum system can evolve
into a classical mixed state, that is, the elements in the reduced density
matrix satisfying 
\begin{eqnarray}
\left\vert \rho _{S}\left( t\right) _{kl}\right\vert &=&0\text{ if }k\neq l,
\notag \\
\left\vert \rho _{S}\left( t\right) _{kl}\right\vert &\neq &0\text{ if }k=l.
\end{eqnarray}

\subsection{Decoherence Without Energy Dissipation}

\begin{figure}[t]
\begin{center}
\includegraphics[width=8.25cm]{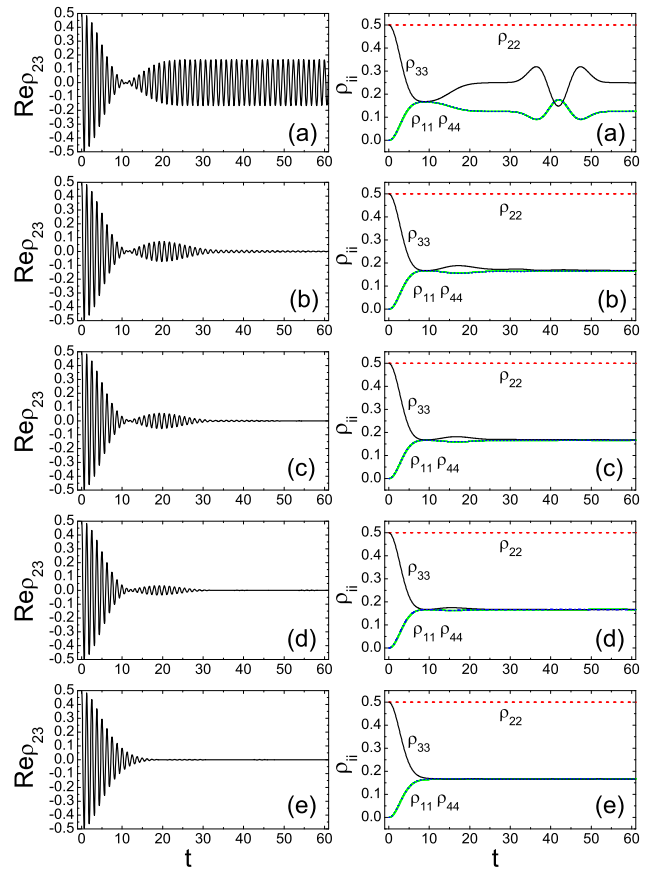}
\end{center}
\caption{Time evolution of the real part of the off-diagonal element $%
\protect\rho _{23}$ (left panel) and the diagonal elements $\protect\rho %
_{11},\ldots ,\protect\rho _{44}$ (right panel) of the reduced density
matrix of a Heisenberg two-spin system ($J=-5$), coupled via an isotropic
Heisenberg interaction $H_{SE}$ ($\Delta =-0.075$ ) to a Heisenberg-type
environment $H_{E}$ ($\Omega =0.1$) with different connectivity: (a) $K=0$;
(b) $K=2$; (c) $K=4$; (d) $K=6$; (e) $K=N-1$. Adapted from Ref. \protect\cite%
{Yuan2008}.}
\label{figferro}
\end{figure}

\begin{figure*}[t]
\begin{center}
\includegraphics[width=16cm]{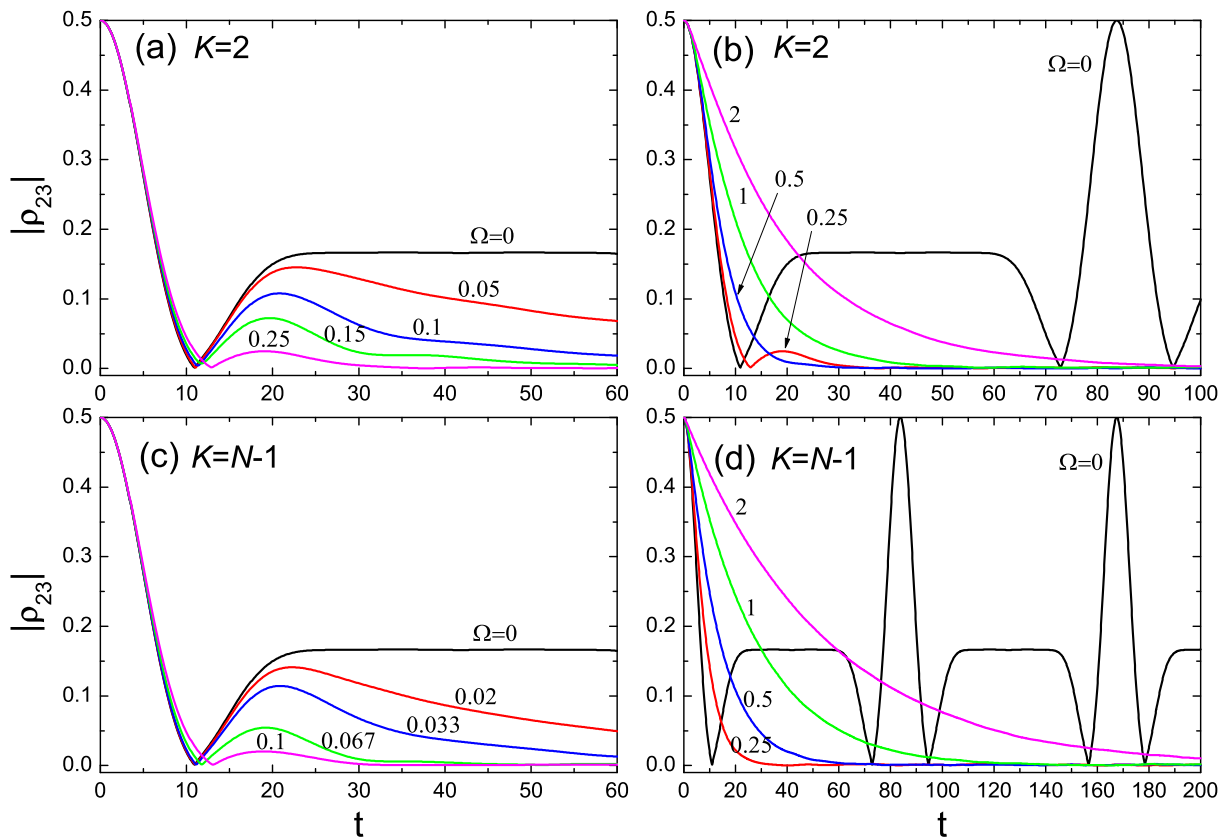}
\end{center}
\caption{Time evolution of the off-diagonal element $\protect\rho _{23}$ of
the reduced density matrix of a Heisenberg two-spin system ($J=-5$),
interacting with a Heisenberg-type environment $H_{E}$ via an isotropic
Heisenberg Hamiltonian $H_{SE}$ ($\Delta =-0.075$ ) for the same geometric
structures in the environment: (a,b) $K=2$ and (c,d) $K=N-1$. The number
next to each curve is the corresponding value of $\Omega $. Adapted from
Ref. \protect\cite{Yuan2008}.}
\label{figheisenberg}
\end{figure*}

In previous work \cite{Yuan2006,Yuan2007,Yuan2008,Yuan2009}, it was shown
that a frustrated environment, such as described by a Heisenberg-type $H_{E}$%
, can enhance the decoherence of the quantum system. The typical results of
the (full) decoherence in the Heisenberg two-spin system without energy
dissipation are the following.

In Fig.~\ref{figferro} and Fig.~\ref{figheisenberg}, we show the time
evolution of the elements of the reduced density matrix $\rho \left(
t\right) $ for different connectivity $K$ with same $\Omega $, or different $%
\Omega $ with same $K$, for the case $[H_{S},H_{SE}]=0$.

If $\left\vert \Delta \right\vert \gg \Omega\sqrt{K} $, in the absence of
interactions between the environment spins ($\Omega\sqrt{K} =0$) and after
the initial decay, the quantum system exhibits long-time oscillations (see
Fig.~\ref{figferro}(a)(left)). As shown in Ref. \cite{ourPRL,Melikidze2004},
in the limit of a large environment ($N\rightarrow \infty $)%
\begin{equation}
\hbox{Re }\rho _{23}\left( t\right) =\left[ \frac{1}{6}+\frac{1-bt^{2}}{3}%
e^{-ct^{2}}\right] \cos \omega t,  \label{melik}
\end{equation}%
where $b=N\Delta ^{2}/4$, $c=b/2$ and $\omega =J-\Delta $. Equation~(\ref%
{melik}) clearly shows the two-step process, that is, after the initial
Gaussian decay of the amplitude of the oscillations, the oscillations revive
and their amplitude levels of by a factor of ~$1/3$ (see Ref. \cite%
{Melikidze2004}). Due to conservation laws, this behavior does not change if
we introduce an isotropic Heisenberg Hamiltonian in the environment ($\Omega
_{i,j}^{(\alpha )}\equiv \Omega $ for all $\alpha $, $i$ and $j$),
independent of $K$. This is also confirmed by our numerical results (not
shown).

If $\left\vert \Delta \right\vert \approx \Omega\sqrt{K} $, the initial
Gaussian decay of the quantum system is not sensitive to the presence of a
Heisenberg-type environment $H_{E}$, but there is a decay of the amplitude
of the long-living oscillations. The larger $K$ (see Fig.~\ref{figferro}%
(b-e)(left)) or $\Omega $ ({see Fig.~\ref{figheisenberg}(a,c)}), the faster
the decay is.

If $\left\vert \Delta \right\vert \ll \Omega\sqrt{K} $ and $\Omega $ is
comparable with $J$, keeping $K$ fixed and increasing $\Omega $ smoothly
changes the initial decay from Gaussian (fast) to exponential (slow). The
long-living oscillations are completely suppressed ({see Fig.~\ref%
{figheisenberg}(b,d)}). For large $\Omega $, the simulation data fits very
well to 
\begin{equation}
\left\vert \rho _{23}\left( t\right) \right\vert =\frac{1}{2}e^{-A_{K}\left(
\Omega \right) t},  \label{p23_exp}
\end{equation}%
where $A_{K}\left( \Omega \right) $ is approximately linearly dependent on $%
\Omega $:%
\begin{equation}
A_{K}\left( \Omega \right) \approx \Omega \widetilde{A}_{K},
\end{equation}%
and we find that $\widetilde{A}_{2}=9.13$ and $\widetilde{A}_{N-1}=26.73$.

\begin{figure*}[t]
\begin{center}
\includegraphics[width=16cm]{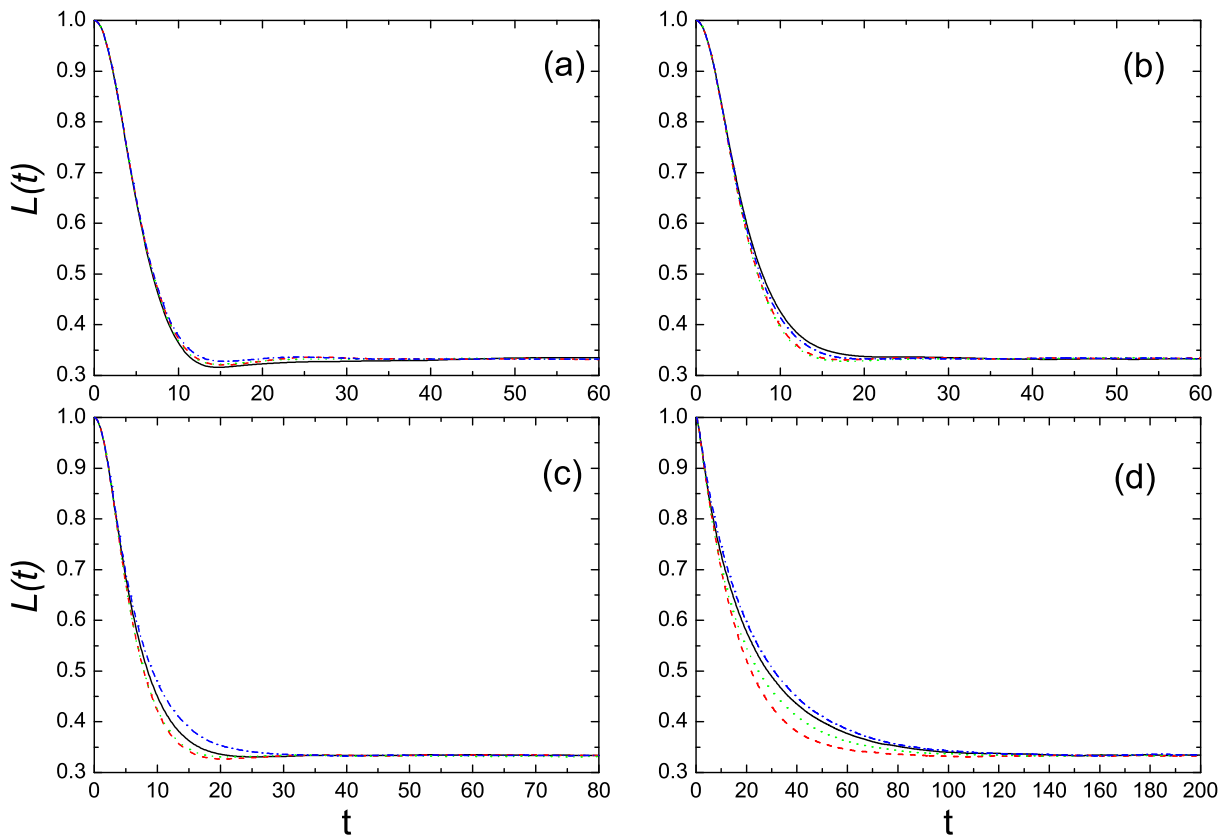}
\end{center}
\caption{Time evolution of the Loschmidt echo $L\left( t\right) $ of a
Heisenberg two-spin system ($J=-5$), interacting with a Heisenberg-type
environment $H_{E}$ via a Heisenberg ($\Delta =-0.075$) Hamiltonian $H_{SE}$%
. The values of $\Omega \protect\sqrt{K}$ are: (a) $\Omega \protect\sqrt{K}%
=0.1\protect\sqrt{N-1}$, (b) $\Omega \protect\sqrt{K}=0.15\protect\sqrt{N-1}$%
, (c) $\Omega \protect\sqrt{K}=0.25\protect\sqrt{N-1}$, and (d) $\Omega 
\protect\sqrt{K}=\protect\sqrt{N-1}$. The different lines in each pannel
correspond to different $K$. Solid (black) line: $K=2$; dashed (red) line: $%
K=4$; dotted (green) line: $K=6$, and dash-dotted (blue) line: $K=N-1$.
Adapted from Ref. \protect\cite{Yuan2008}.}
\label{figtwo5}
\end{figure*}

Physically, the observed behavior can be understood as follows \cite%
{Yuan2008}. If $\left\vert \Delta \right\vert \approx \Omega\sqrt{K} $, a
bath spin is roughly equally affected by the motion of the other bath spins
and the system spins. Therefore, each bath spin follows the original
dynamics, as if there was no coupling between bath spins. This explains why
the initial Gaussian decay is insensitive to the values of $K$ or $\Omega $.
After the initial decay, the whole system is expected to reach a stationary
state, but because of the presence of Heisenberg-type interactions between
the bath spins, it leads to a decrease of the coherence between the singlet
and triplet states, and therefore a new stationary state of the bath is
established, suppressing the long-living oscillations.

For larger $K$, the distance between two bath spins, defined as the minimum
number of bonds connecting the two spins, becomes smaller. For instance, for 
$K=2$, this distance is $\left( N-2\right) /2$, and for $K=N-1$, it is zero.
For fixed $\Omega $ and larger $K$ the fluctuations in the spin bath
propagate faster, and therefore the evolution to the stationary state is
faster. Furthermore, since the environment in our model is a highly
frustrated system, increasing the connectivity $K$ will increase the energy
resolution of the eigenstates, which makes the dynamics of the environment
more complicated. For fixed $K$, increasing the coupling strength between
the bath spins will speed up the dynamics of the bath, that is, the larger $%
\Omega $ the faster will be the evolution to the stationary state. However
the coupling strength within the environment should not be too large,
because otherwise the energy resolution in the bath will be too small to
lead the energy dissipation of the quantum system.

In the opposite case $\left\vert \Delta \right\vert \ll \Omega\sqrt{K} $ and 
$\Omega $ is comparable with $J$, $H_{SE}$ is a small perturbation relative
to $H_{E}$ and the coupling between the bath spins is the dominant factor in
determining the dynamics of the bath spins. Therefore, by increasing $K$ or $%
\Omega $, the bath spins will have less freedom to follow the dynamics
induced by the coupling to the two system spins, the influence of the bath
on the quantum system will decrease, and the (exponential) decay will become
slower.

Here we have compared $\Omega\sqrt{K} $ to $\left\vert \Delta \right\vert $
to distinguish different regimes. As a matter of fact, $\Omega\sqrt{K} $
does not completely characterize the decoherence process, but it can be used
to characterize its time scale. Indeed, as shown in Fig.~\ref{figtwo5}, for
different $\sqrt{K}$ and $\Omega $ but the same value of $\Omega\sqrt{K} $,
the time evolution of $L(t)$ is very similar. Note that if $\Omega\sqrt{K} $
increases (compare Fig.~\ref{figtwo5}a to Fig.~\ref{figtwo5}d), the
differences between the Loschmidt echoes increase.

According to the general picture of decoherence~\cite{Zurek2003}, for an
environment with nontrivial internal dynamics that is initially in a random
superposition of all its eigenstates, we expect that the quantum system will
evolve into a stable mixture of its eigenstates. In other words, the
decoherence will cause all the off-diagonal elements of the reduced density
matrix to vanish with time. In the case of an isotropic Heisenberg coupling
between the quantum system and the environment, $H_{S}$ commutes with the
Hamiltonian $H$, hence the energy of the quantum system is a conserved
quantity. Therefore, the weight of the singlet $\left\vert {S}\right\rangle $
in the mixed state should be a constant ($1/2$), and the weights of the
degenerate eigenstates $|T_{0}\rangle $, $|T_{-1}\rangle $ and $%
|T_{1}\rangle $ are expected to become the same (${1/6}$). As shown in Fig.~%
\ref{figferro}(b-e)(right), our simulations confirm that this picture is
correct in all respects.

\subsection{Decoherence With Energy Dissipation}

\begin{figure}[t]
\begin{center}
\includegraphics[width=8.25cm]{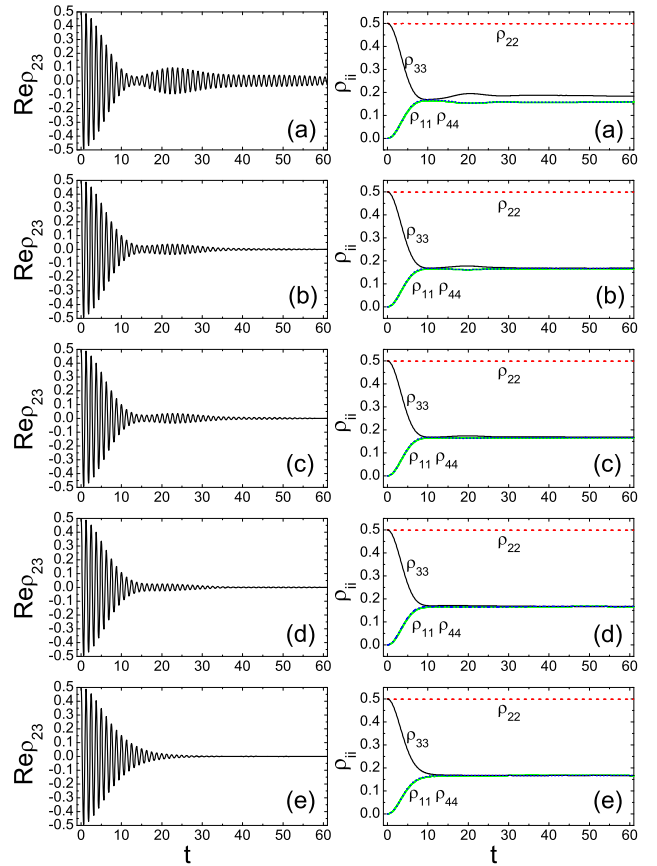}
\end{center}
\caption{Time evolution of the real part of the off-diagonal element $%
\protect\rho _{23}$ (left panel) and the diagonal elements $\protect\rho %
_{11},\ldots ,\protect\rho _{44}$ (right panel) of the reduced density
matrix of a Heisenberg two-spin system ($J=-5$), coupled via an isotropic
Heisenberg-type interaction $H_{SE}$ ($\Delta =-0.15$ ) to a Heisenberg-type
environment $H_{E}$ ($\Omega =0.15$) with different connectivity: (a) $K=0$;
(b) $K=2$; (c) $K=4$; (d) $K=6$; (e) $K=N-1$. }
\label{two_ed}
\end{figure}

\begin{figure*}[t]
\begin{center}
\includegraphics[width=16cm]{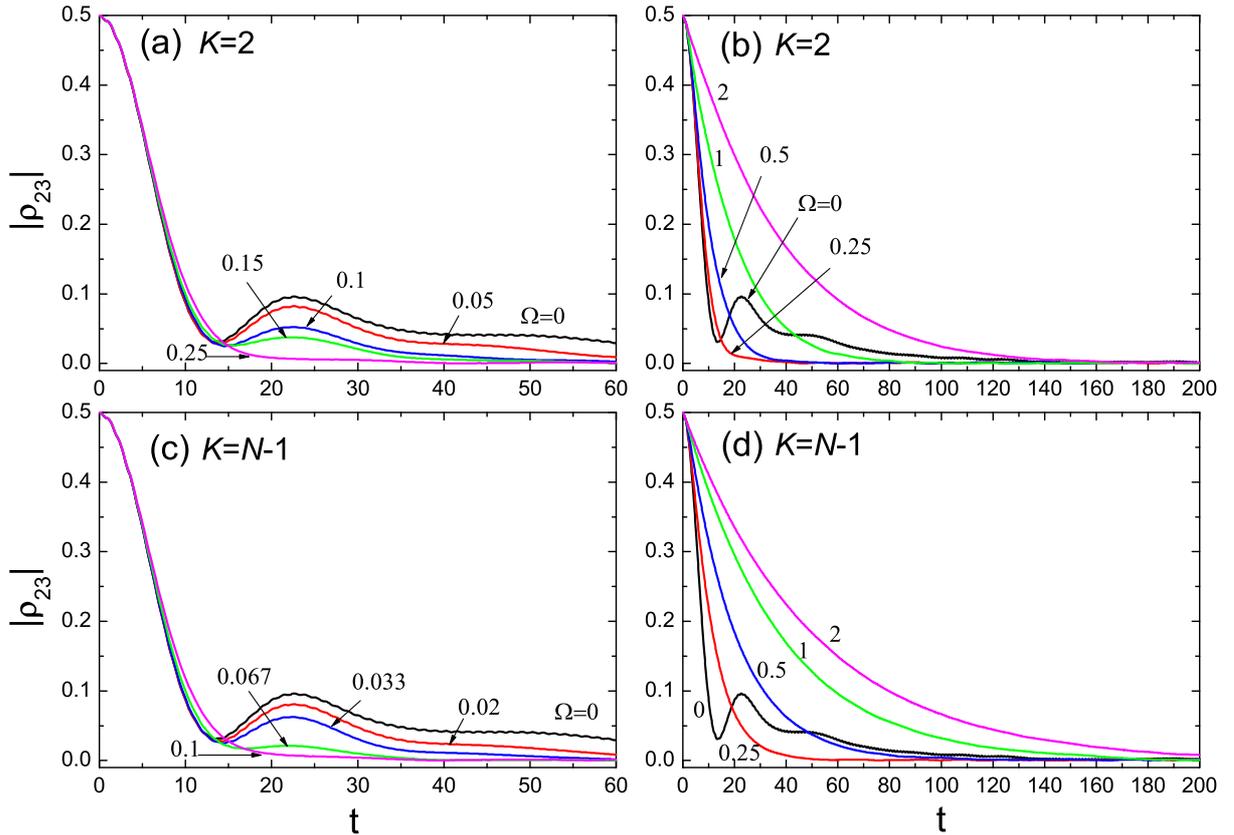}
\end{center}
\caption{Same as Fig.~\protect\ref{figheisenberg} except that $H_{SE}$ is
Heisenberg-type and $\Delta =0.15$. Adapted from Ref. \protect\cite{Yuan2008}%
.}
\label{two_p}
\end{figure*}

Now we consider the case that there is energy dissipation of the quantum
system, i.e. $[H_{S},H]\neq 0$. First, instead of considering a Heisenberg
system-environment interaction $H_{SE}$ as in Fig.~\ref{figferro}-\ref%
{figheisenberg}, we now take a Heisenberg-type $H_{SE}$ in Fig.~\ref{two_ed}-%
\ref{two_p}, the other interactions like $H_{S}$ and $H_{E}$ are the same as
in Fig.~\ref{figferro}-\ref{figheisenberg}. From a direct comparison of
these results, it is clear that the roles of $K$ and $\Omega $ are the same,
no matter whether the energy of the quantum system is conserved or not. If $%
\left\vert \Delta \right\vert \gg \Omega \sqrt{K}$, in the presence of
anisotropic interactions between the quantum system and the environment
spins, the second step of the oscillations decay and finally disappear as $K$
increases, even in the absence of interactions between the bath spins. This
is because the anisotropic interactions break the rotational symmetry of the
coupling between the quantum system and the environment which is required
for the long-living oscillations to persist. If $\left\vert \Delta
\right\vert \ll \Omega \sqrt{K}$ and $\Omega $ is comparable with $J$, $%
\left\vert \rho _{23}\left( t\right) \right\vert $ can still be described by
Eq.~(\ref{p23_exp}), but now $A_{K}\left( \Omega \right) $ is no longer a
linear function of $\Omega $. This is because the energy dissipation will
change the weight of each pointer state (eigenstate) in the final stable
mixture, which makes the time evolution of $\left\vert \rho _{23}\left(
t\right) \right\vert $ more complicated.

\begin{figure*}[t]
\begin{center}
\mbox{
\includegraphics[width=8.25cm]{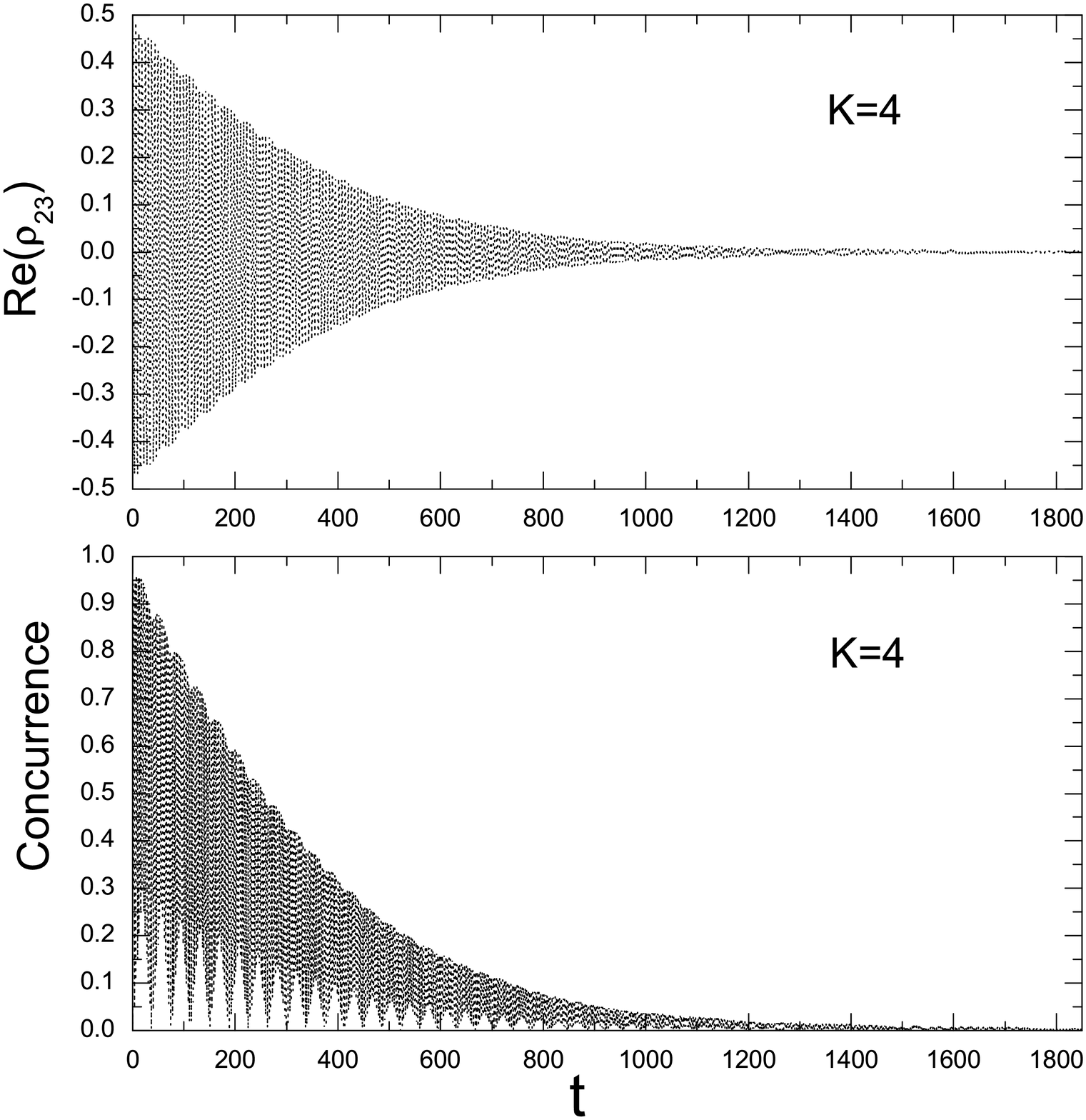}
\includegraphics[width=8.25cm]{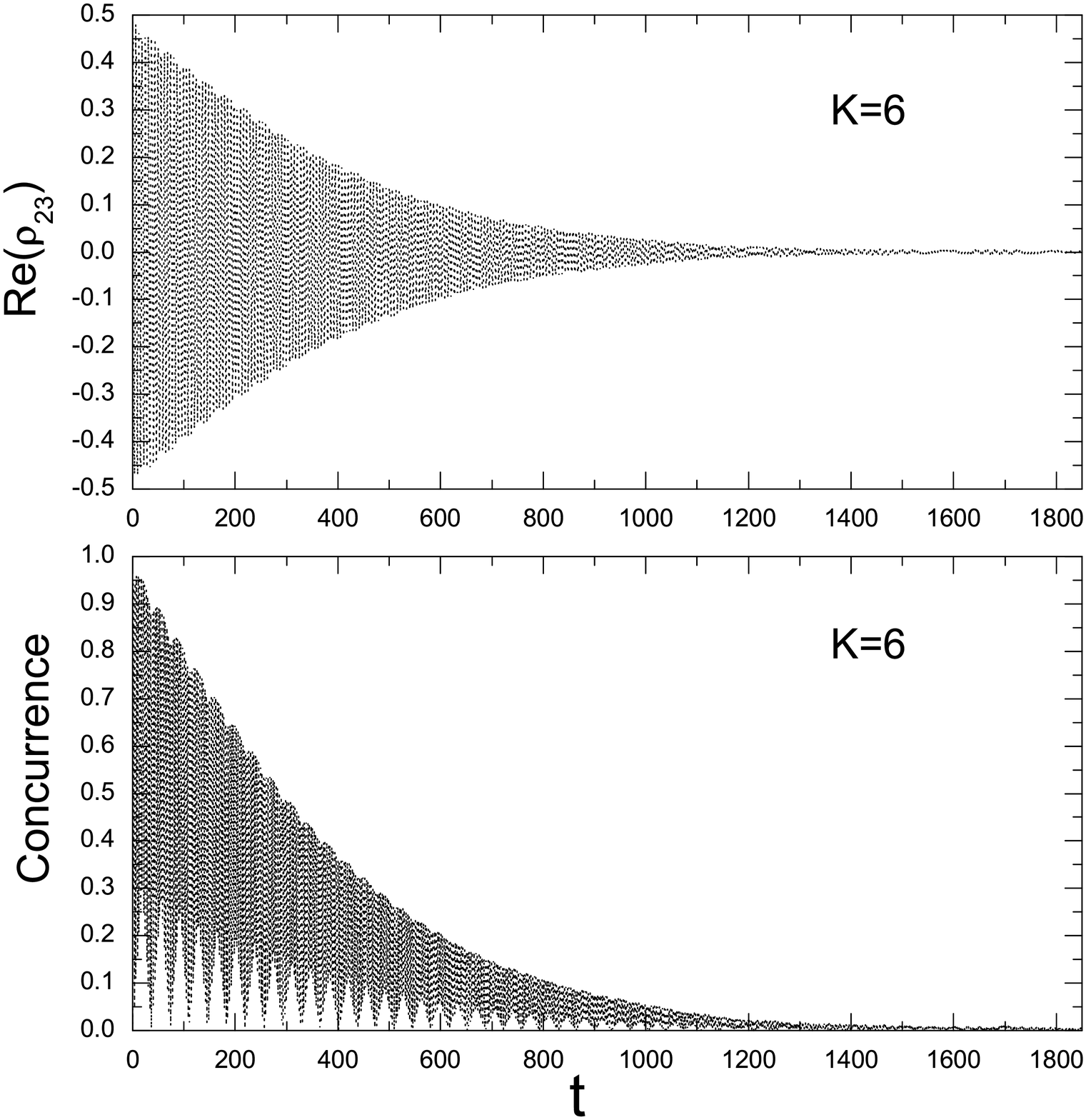}
}
\end{center}
\caption{Time evolution of the real part of the off-diagonal element $%
\protect\rho _{23}$ and concurrence of the reduced density matrix of a
Heisenberg two-spin system ($J=-1$), coupled via an Ising interaction $\pm $ 
$H_{SE}$ ($\Delta =0.075$ ) to an isotropic Heisenberg environment $H_{E}$ ($%
\Omega =-0.15$, $N=16$) with connectivity $K=4$ (square lattice) and $K=6$
(triangle lattice).}
\label{isingst18}
\end{figure*}

More results with Ising$\pm $ interaction $H_{SE}$ are shown in Fig.~\ref%
{ringjedifferent} and Fig.~\ref{isingst18}. They give similar results as in
the case of a Heisenberg-type interaction $H_{SE}$. The environments in Fig.~%
\ref{isingst18} are not a spin glass, but an isotropic Heisenberg
antiferromagnetic square or triangle lattice, which are also frustrated
systems.

\begin{figure}[t]
\begin{center}
\includegraphics[width=8.25cm]{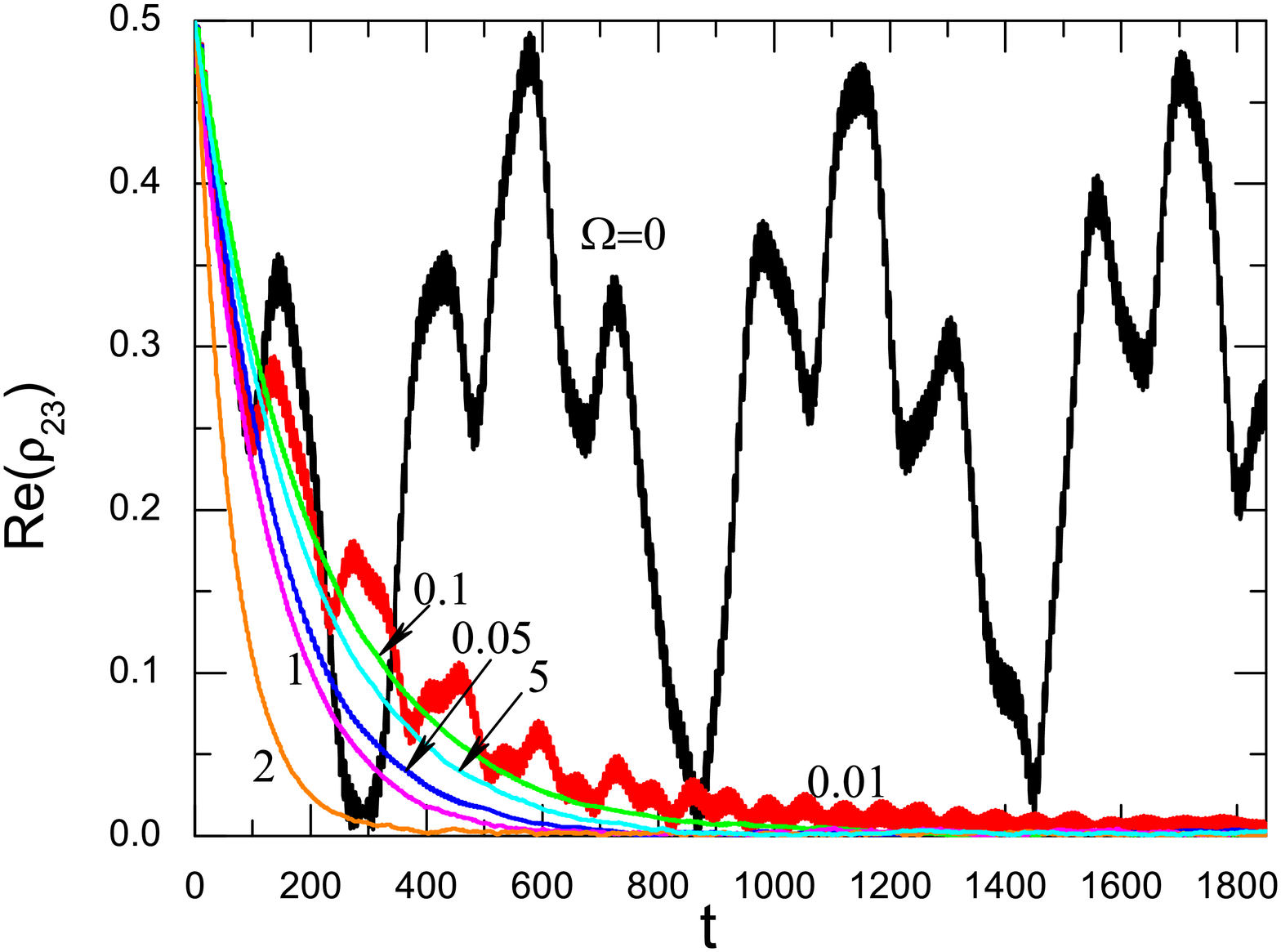}
\end{center}
\caption{Time evolution of the real part of the off-diagonal element $%
\protect\rho _{23}$ of the reduced density matrix of a Heisenberg two-spin
system ($J=-5$), coupled via an Ising$\pm $ interaction $H_{SE}$ ($\Delta
=0.075$ ) to a Heisenberg-type-ring environment $H_{E}$ ($N=16$) with
different range of $\Omega $ ($0,0.01,0.05,0.1,1,2,5)$. The value of the
corresponding $\Omega $\ is indicated by the number near each line.}
\label{ringjedifferent}
\end{figure}

\subsection{Summary}

In conclusion, with a frustrated spin-bath environment that initially is in
a random superposition of its basis states, a pure quantum state of the
quantum spin system will evolve into a classical mixed state. If the
interaction between the quantum system and environment is much smaller than
the coupling between the spins in the quantum system, the pointer states are
the eigenstates of the quantum system. Both of these observations are in
concert with the general picture of decoherence~\cite{Zurek2003}.
Furthermore, if the energy of the quantum system is conserved, the pointer
states can still be the eigenstates of the quantum system, independent of
the ratio of the system-bath coupling to the coupling within the system. For
the anisotropic spin-bath, changing the internal dynamics of the environment
(geometric structure or exchange couplings) may change the decoherence of
the quantum spin system from Gaussian to exponential decay.

\section{Evolution to the Ground State of a Two-spin System}

To approach the ground state of a quantum system by coupling it to a large
quantum bath is not a trivial problem. The interactions between the two
quantum systems will not only lead to the exchange of the energy, but also
the coherence of the wave function. In this section, we show under which
condition a two-spin system can approach its ground or near-ground state.
The affect of entanglement and interaction symmetry during this evolution
will be discussed.

The interaction within the two-spin system will be fixed as isotropic
Heisenberg ferromagnet ($J>0$) or antiferromagnet ($J<0$). In the case of
ferromagnet, there are three degenerate ground states (triplet states) $%
\left\vert T_{0}\right\rangle $, $\left\vert T_{1}\right\rangle $ and $%
\left\vert T_{-1}\right\rangle $, whereas in the case of antiferromagnet it
is the singlet state $\left\vert S\right\rangle $. Therefore when the
quantum system approaches the state which has the same energy as the ground
state(s), it can be a single eigenstate if the ground state is
non-degenerate, and it can also be an entangled superposition (quantum) or
mixed state (classic) if the ground states are degenerate.

In order to let the two-spin system approach the ground state, it is
necessary to keep the environment at low temperature. Instead of a classical
mixture, we will prepare the environment in a pure quantum state, in which
the temperature is not well defined. So we simply initialize the environment
in its ground or near-ground state to guarantee the one-direction energy
flow. As in Ref. \cite{Yuan2008,Yuan2009}, this initial ground state of the
environment leads to a sharp local density of states in the whole system,
and therefore the decoherence of the quantum system is much weaker comparing
to cases with random initial state of the environment.

The interactions between the two-spin system and the spin bath will be set
as Heisenberg-type or Ising-type. Even in both cases the interactions of
different orientations (x,y and z) are not all the same, but they are
totally different: the Heisenberg-type $H_{SE}\,$\ is still symmetrical
because the exchange interactions have the same random amplitude in the
three orientations, whereas the Ising-type $H_{SE}$ is antisymmetric since
the exchange interactions are totally different between x(y) and z
directions.

The interactions within the environment are fixed as Heisenberg-type, which
reduce the strength of the coherence between the quantum system and the
environment, as we showed in the previous section and Ref. \cite%
{Yuan2006,Yuan2007,Yuan2008,Yuan2009}.

\subsection{Symmetrical Coupling}

\begin{figure}[t]
\begin{center}
\includegraphics[width=8.25cm]{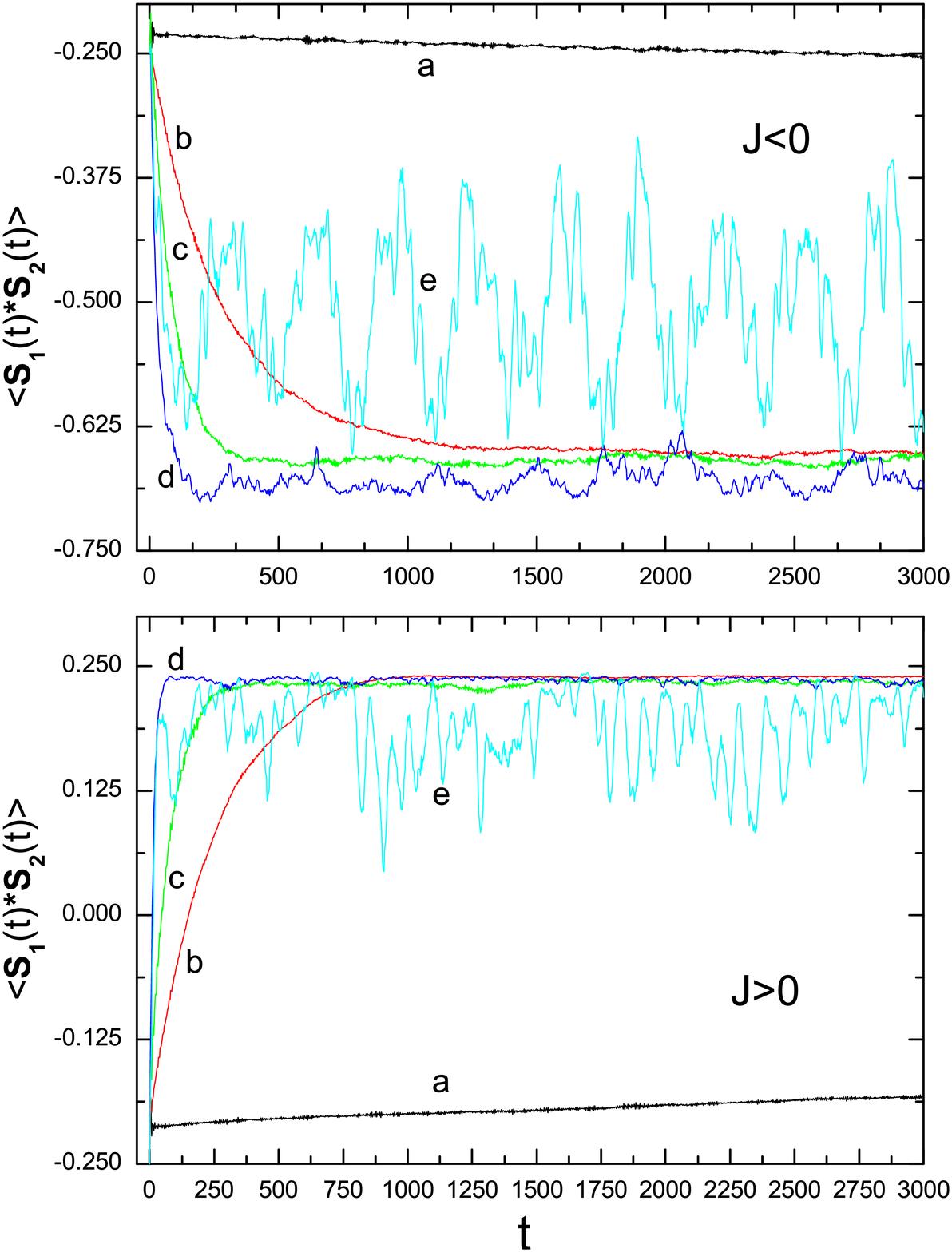}
\end{center}
\caption{Time evolution of the correlation $\langle \Psi (t)|\mathbf{S}%
_{1}\cdot \mathbf{S}_{2}|\Psi (t)\rangle $ of the antiferromagnetic (top
panel) and ferromagnetic (bottom panel) two-spin quantum system with
Heisenberg-type interaction $H_{SE}$ and environment $H_{E}$. The model
parameters are $\Delta =0.15$ and a: $\Omega =0.075$; b: $\Omega =0.15$; c: $%
\Omega =0.20$; d: $\Omega =0.30$; e: $\Omega =1$. The number of spins in the
environment is $N=14$.}
\label{compareje16}
\end{figure}

\begin{figure}[t]
\begin{center}
\includegraphics[width=8.25cm]{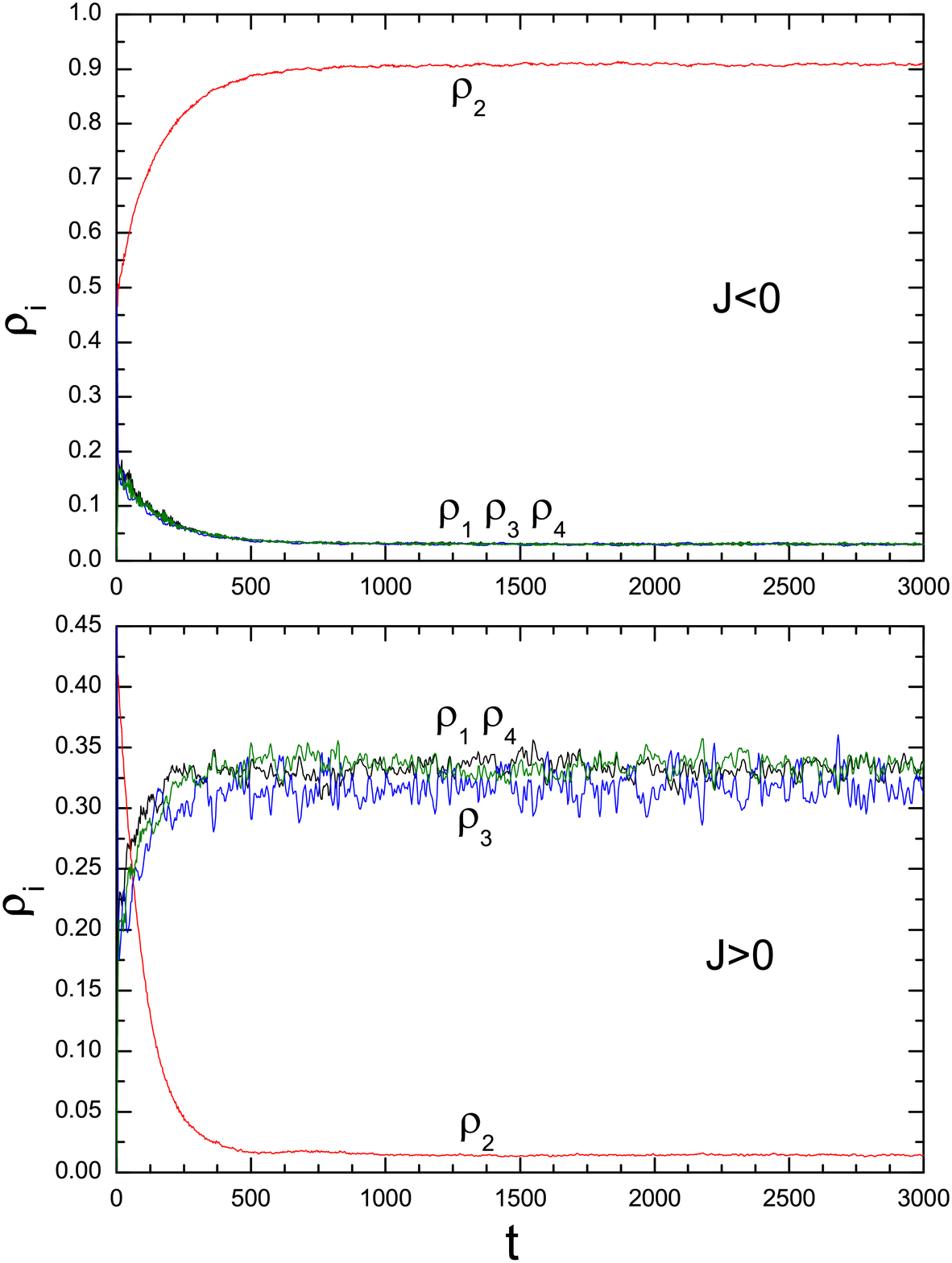}
\end{center}
\caption{Time evolution of the diagonal matrix elements of the reduced
density matrix of the antiferromagnetic (top panel) and ferromagnetic
(bottom panel) two-spin quantum system for $\Delta =0.15$ and $\Omega =0.15$
(case (b) of Fig.~\protect\ref{compareje16}, except that the number of spins
in the environment is $N=16$)}
\label{hseherdensity18}
\end{figure}

We first consider the case that an antiferromagnetic ($J<0$) or
ferromagnetic ($J>0$) quantum system that interacts with the Heisenberg-type
environment via a Heisenberg-type interaction.

In Fig.~\ref{compareje16}, we present simulation results for the two-spin
correlation function (as a measure of the energy) for different values of
the coupling strength ($\Omega )$ in the environment. Clearly, in case (a),
the relaxation in both cases, antiferromagnetic and ferromagnetic, is rather
slow and confirming that there is relaxation to the ground state requires a
prohibitively long simulation. For cases (b) -- (d), the results are in
concert with the intuitive picture of relaxation due to decoherence: The
correlation shows the relaxation from the up-down initial state of the
quantum system to the ground or near-ground state. As for the
antiferromagnetic quantum system ($J<0$), the two-spin correlation relaxes
to a value of about $0.65$ -- $0.70$, which is much further away from the
ground state value $-3/4$ than we would have expected on the basis of the
results of the ferromagnetic quantum system. In the true ground state of the
whole system, the value of the two-spin correlation in case (b) of $J<0$ is $%
-0.7232$, and hence significantly lower than the typical values, reached
after relaxation. On the one hand, it is clear (and to be expected) that the
coupling to the environment changes the ground state of the quantum system,
but on the other hand, our numerical calculations show that this change is
too little to explain the apparent difference from the results obtained from
the time-dependent solution. In case (e), the characteristic strength of the
interactions between the spins in the environment is of the same order as
the exchange coupling in the quantum system ($\Omega \approx J$), a regime
in which there clearly is significant transfer of energy, back-and-forth,
between the quantum system and the environment.

In Fig.~\ref{hseherdensity18}, we show the diagonal elements of the reduced
density matrix for case (b). After reaching the steady state, the
nondiagonal elements exhibit minimum fluctuations about zero and are
therefore not shown. From Fig. \ref{hseherdensity18}, it is then clear that
the quantum system relaxes to the singlet state for the antiferromagnetic ($%
J<0$) system, and to a mixed state for the ferromagnetic system ($J>0$), as
expected on intuitive grounds.

An important observation is that our data convincingly shows that it is not
necessary to have a macroscopically large environment for decoherence to
cause relaxation to the ground state: A spin-glass with $N=14$ spins seems
to be more than enough to mimic such an environment for a two-spin system.
This observation is essential for numerical simulations of relatively small
systems to yield the correct qualitative behavior.

Qualitative arguments for the high efficiency of the spin-glass bath were
given in Ref.~\onlinecite{Yuan2006}. Since spin-glasses possess a huge
amount of states that have an energy close to the ground state energy but
have wave functions that are very different from the ground state, the
orthogonality catastrophe, blocking the quantum interference in the quantum
system~\cite{Zeh1996,Zurek2003} is very strongly pronounced in this case.

From the data for (b) -- (d), shown in Fig.~\ref{compareje16}, we conclude
that the time required to let the quantum system relax to a state that is
close to the ground state depends on the energy scale ($\Omega $) of the
random interactions between the spins in the environment. As it is difficult
to define the point in time at which the quantum system has reached its
stationary state, we have not made an attempt to characterize the dependence
of the relaxation time on $\Omega $.

\subsection{Antisymmetric Coupling}

\begin{figure}[t]
\begin{center}
\includegraphics[width=8.25cm]{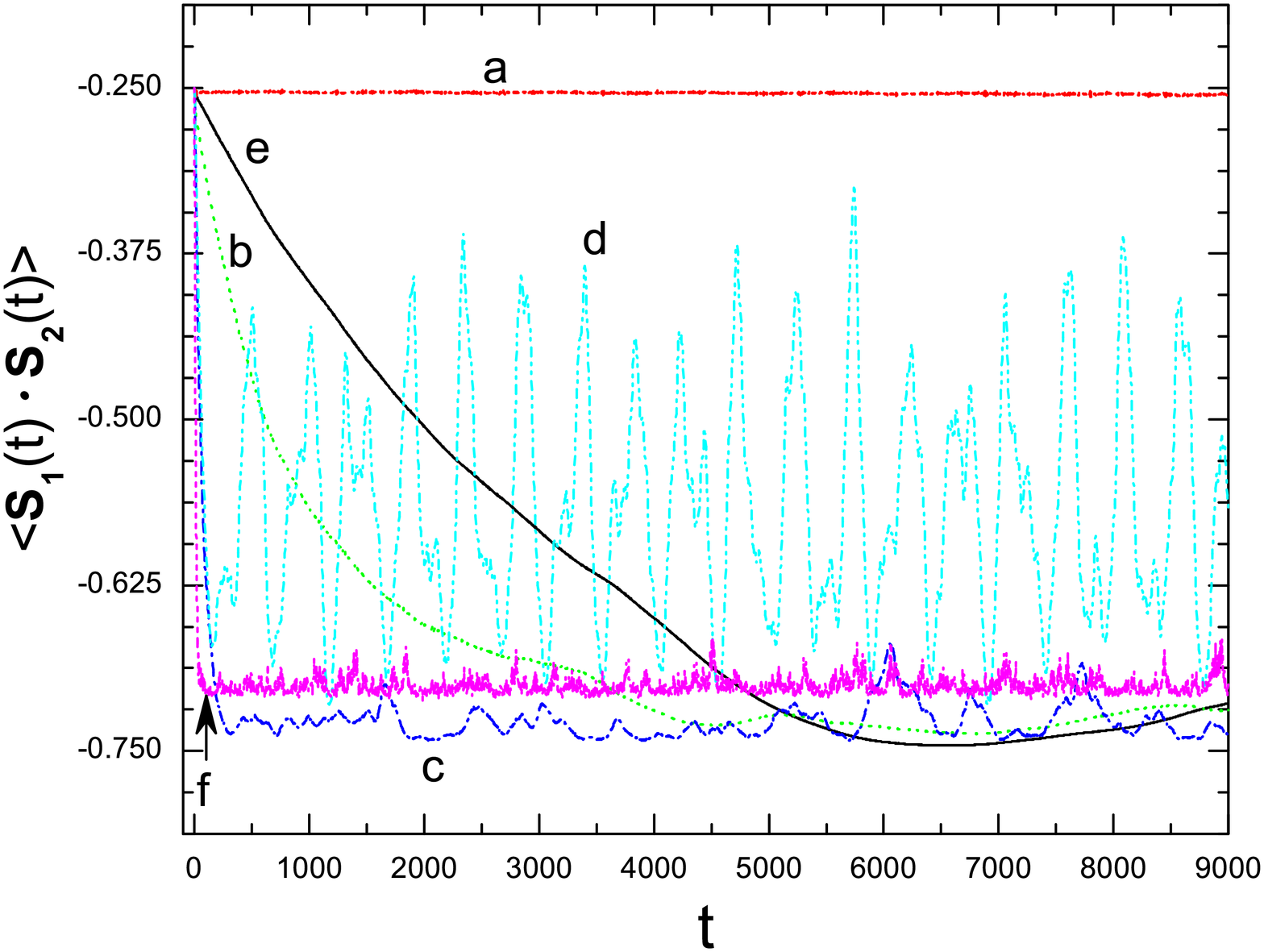}
\end{center}
\caption{Time evolution of the correlation $\langle \Psi (t)|\mathbf{S}%
_{1}\cdot \mathbf{S}_{2}|\Psi (t)\rangle $ of the antiferromagnetic quantum
system with Ising-type $H_{SE}$ and Heisenberg-type $H_{E}$. The model
parameters are (a-d) $\Delta =0.075$, (e) $\Delta =0.0375$, (f) $\Delta
=0.15 $, and (a) $\Omega =0.075$, (b,e) $\Omega =0.15$, (c,f) $\Omega =0.30$%
, (d) $\Omega =1$. The number of spins in the environment is $N=16$.}
\label{figantiferro}
\end{figure}

\begin{figure*}[t]
\begin{center}
\includegraphics[width=16cm]{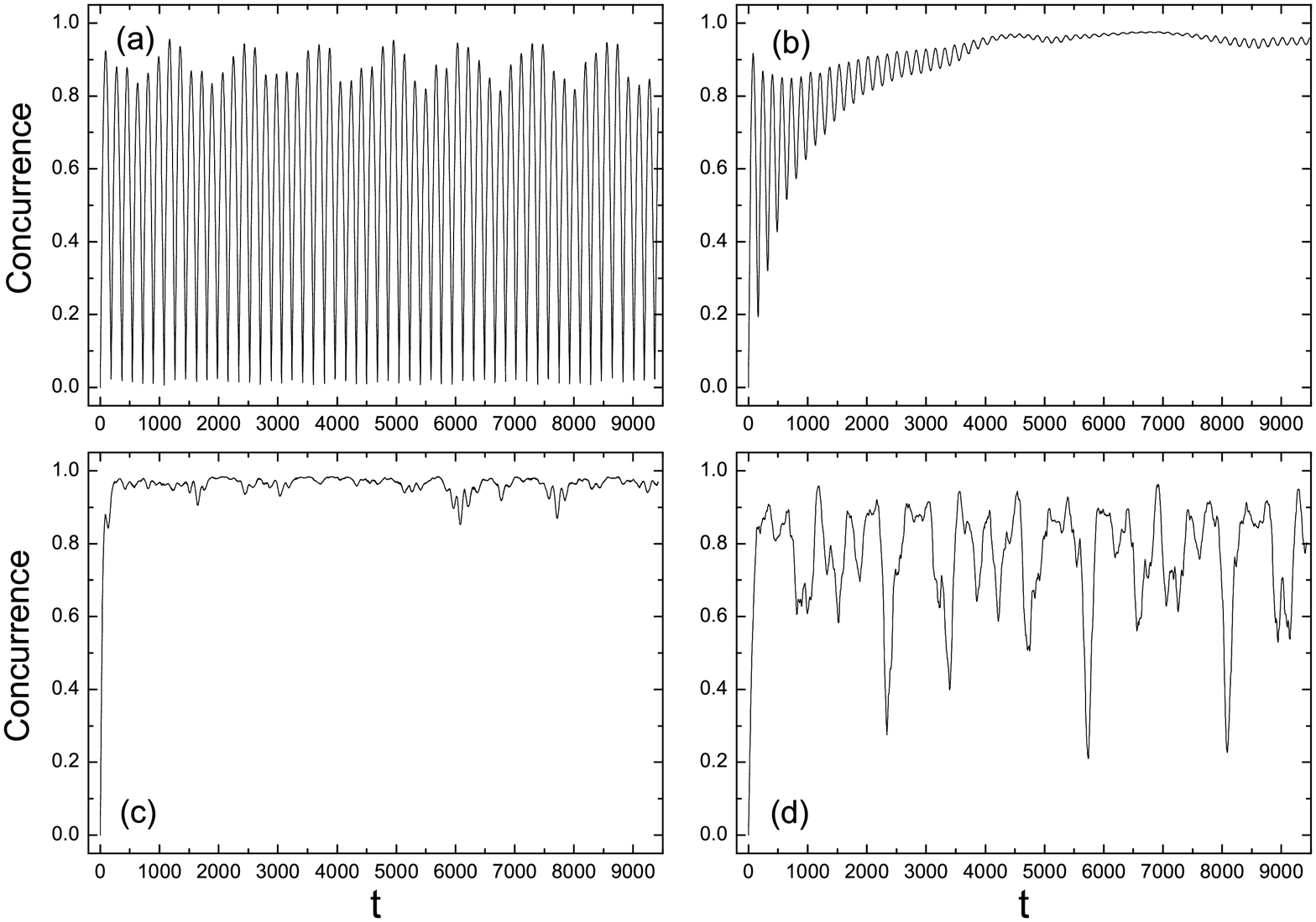}
\end{center}
\caption{Time evolution of the concurrence of the antiferromagnetic quantum
system with Ising-type interaction $H_{SE}$ and Heisenberg-type environment $%
H_{E}$. The model parameters are $\Delta =0.075$ and (a) $\Omega =0.075$,
(b) $\Omega =0.15$, (c) $\Omega =0.30$, (d) $\Omega =1$. The number of spins
in the environment is $N=16$.}
\label{figantiferroconcurrence}
\end{figure*}

In our simulation, the initial state of the quantum system is $\left\vert
\uparrow \downarrow \right\rangle $ and this state has total magnetization $%
M=0$. For an Ising-type interaction $H_{SE}$ of the envionment with a
Heisenberg system $H_{S}$, the magnetization $M$ of the quantum system
commutes with the Hamiltonian of the whole system. Therefore, the
magnetization of the quantum system is conserved during the time evolution,
and the quantum system will always stay in the subspace with $M=0$. In this
subspace, the ground state for the antiferromagnetic quantum system is the
singlet state $|S\rangle $ while for the ferromagnetic quantum system the
ground state (in the $M=0$ subspace) is the entangled state $|T_{0}\rangle $%
. Thus, for the Ising-type interaction $H_{SE}$, starting from the initial
state $\left\vert \uparrow \downarrow \right\rangle $, the quantum system
should relax to an entangled state, for both a ferro- or antiferromagnetic
quantum system, that is, at any time $t$, the state of the whole system can
be written as

\begin{equation}
|\Psi (t)\rangle =|S\rangle |\Phi _{S}(t)\rangle +|T_{0}\rangle |\Phi
_{T_{0}}(t)\rangle ,  \label{Phia}
\end{equation}%
where $|\Phi _{S}\rangle $ and $|\Phi _{T_{0}}\rangle $ denote the states of
the environment.

Let us denote by $\{|\phi _{i}\rangle \}$ a complete set of states of the
environment. Within the subspace spanned by the states $\{|S\rangle |\phi
_{i}\rangle ,|T_{0}\rangle |\phi _{i}\rangle \}$, the Hamiltonian can be
written as 
\begin{eqnarray}
H &=&E_{S}|S\rangle \langle S|+E_{T}|T_{0}\rangle \langle T_{0}|+H_{E} 
\notag \\
&&-\frac{1}{2}\sum_{j=1}^{N}(\Delta _{1,j}^{(z)}-\Delta _{2,j}^{(z)})\left(
|S\rangle \langle T_{0}|+|T_{0}\rangle \langle S|\right) I_{j}^{z},  \notag
\label{HST1} \\
&&
\end{eqnarray}%
where we used $\langle S|S_{1}^{z}|S\rangle =\langle
T_{0}|S_{1}^{z}|T_{0}\rangle =\langle S|S_{2}^{z}|S\rangle =\langle
T_{0}|S_{2}^{z}|T_{0}\rangle =0$, $\langle T_{0}|S_{1}^{z}|S\rangle =1/2$,
and $\langle T_{0}|S_{2}^{z}|S\rangle =-1/2$.

Introducing a pseudo-spin $\sigma =(\sigma ^{x},\sigma ^{y},\sigma ^{z})$
such that the eigenvalues $+1$ and $-1$ of $\sigma ^{z}$ correspond to the
states $|S\rangle $ and $|T_{0}\rangle $, respectively, Eq.~(\ref{HST1}) can
be written as 
\begin{eqnarray}
H &=&\frac{E_{S}-E_{T}}{2}+\frac{E_{S}+E_{T}}{2}\sigma ^{z}+H_{E}  \notag \\
&&-\frac{1}{2}\sum_{j=1}^{N}(\Delta _{1,j}^{(z)}-\Delta
_{2,j}^{(z)})I_{j}^{z}\sigma ^{x},  \label{HST2}
\end{eqnarray}%
showing that in the case of Ising-type $H_{SE}$, the quantum system with two
spins is equivalent to the model Eq.~(\ref{HST2}) with one spin.

From Eq.~(\ref{HST2}), it follows immediately that the Hamiltonian is
invariant under the transformation $\{J,\sigma ^{z}\}\rightarrow
\{-J,-\sigma ^{z}\}$. Indeed, the first, constant term in Eq.~(\ref{HST2})
is irrelevant and we can change the sign of the second term by rotating the
speudo-spin by 180 degrees about the $x$-axis. Therefore, if the initial
state is also invariant under this transformation, the time-dependent
physical properties will not depend on the choice of the sign of $J$, hence
the ferro- and antiferromagnetic system will behave in exactly the same
manner. In our case, the initial state can be written as $(|S\rangle
+|T_{0}\rangle )|\phi _{0}\rangle /\sqrt{2}$, which is trivially invariant
under the transformation $\sigma ^{z}\rightarrow -\sigma ^{z}$.

Therefore for Ising-type $H_{SE}$ ($\Delta _{i,j}^{(x)}=\Delta
_{i,j}^{(y)}=0 $), an initial state that is invariant for the transformation 
$|S\rangle \leftrightarrow |T_{0}\rangle )$, $\langle \Psi (t)|A|\Psi
(t)\rangle $ does not depend on the sign of $J$, for any observable $A$ of
the quantum system that is invariant for this transformation. Under these
conditions, it is easy to prove that 
\begin{equation}
\langle \Psi (t)|\mathbf{S}_{1}\cdot \mathbf{S}_{2}|\Psi (t)\rangle
_{F}+\langle \Psi (t)|\mathbf{S}_{1}\cdot \mathbf{S}_{2}|\Psi (t)\rangle
_{A}=-\frac{1}{2},  \label{proof1}
\end{equation}%
where the subscript $F$ and $A$ refer to the ferro- and antiferromagnetic
quantum system, respectively.

Likewise, for the concurrence we find $C_{F}\left( t\right) =C_{A}\left(
t\right) $ and similar symmetry relations hold for the other quantities of
interest. Of course, this symmetry is reflected in our numerical data also,
hence we can limit ourselves to present data for the antiferromagnetic
quantum system with Ising-type $H_{SE}$.

In Fig.~\ref{figantiferro} and \ref{figantiferroconcurrence}, we present
simulation results for the two-spin correlation function and concurrence for
different values of the parameter $\Delta $ and $\Omega $. It is clear that
for a certain range of the interaction strength, the quantum system relaxes
to a state that is very close to the ground state, see Fig.~\ref%
{figantiferro}(c) and \ref{figantiferroconcurrence}(c). That is, the
presence of a conserved quantity (the magnetization of the quantum system)
acts as a catalyzer for relaxing to the ground state. Intuitively, we would
expect that the presence of a conserved quantity hinders the relaxation and
indeed, the relaxation in Fig.~\ref{figantiferro} is much slower than in
Fig.~\ref{compareje16}. Notwithstanding this, in the presence of a conserved
quantity, the quantum system relaxes to a state that is much closer to the
true ground state than the one it would relax to in the absence of this
conserved quantity.

One should notice that only in a small range of parameters $\Delta $ and $%
\Omega $ the two-spin quantum system can evolve into a near-ground state. If
the interactions within the bath is too small, the range of the energy
spectrum of the bath limits the energy dissipation of the system, and the
system spins mainly follow its own dynamics just like there is no bath. On
the contrary, if the interactions within the bath are too strong, a small
change of the configuration of the bath spins will lead to a large change of
its energy, therefore the direction of the energy flow will oscillate with
time and the quantum system cannot arrive to a state with approximately
stable energy.

\subsection{Summary}

In general, it turns out that the relaxation to the ground state is a more
complicated process than one would naively expect, depending essentially on
the ratio between parameters of the interaction and environment
Hamiltonians. Two general conclusions are:

(1) the quantum system more easily evolves into its ground state when the
latter is more degenerate (degenerate triplet states compared to the
singlet) or less entangled (e.g. up-down state compared to the singlet);

(2) constraints on the system such as existence of additional integrals of
motion can make the evolution to the ground state more efficient.

An explanation of the first statement is that if the ground state is more
degenerate, its affective dimension in the Hilbert space is larger and
therefore the wave function will have more possibilities to evolve into this
subspace, especially if the environment is frustrated. The latter statement
looks a bit counterintuitive since it means that it may happen that a more
regular system exhibits stronger relaxation than a chaotic one. The reason
that it may happen is that introducing an additional integral of motion,
such as the total magnetization, limits the dimensionality of the available
Hilbert space for the quantum system. The larger the dimensionality of the
available Hilbert space, the more complicated the decoherence process is due
to the appearance of the whole hierarchy of decoherence. A manifestation of
this phenomenon has been observed earlier~\cite{ourPRL}: Under certain
conditions, the same quantum system as studied here (four by four reduced
density matrix) displays \textquotedblleft quantum oscillations without
quantum coherence\textquotedblright\ whereas for a single spin in a magnetic
field (two by two reduced density matrix) decoherence can, relatively
easily, suppress the Rabi oscillations completely.

\section{Decoherence of a Many-Spin System}

\begin{figure*}[t]
\begin{center}
\mbox{
\includegraphics[width=8.25cm]{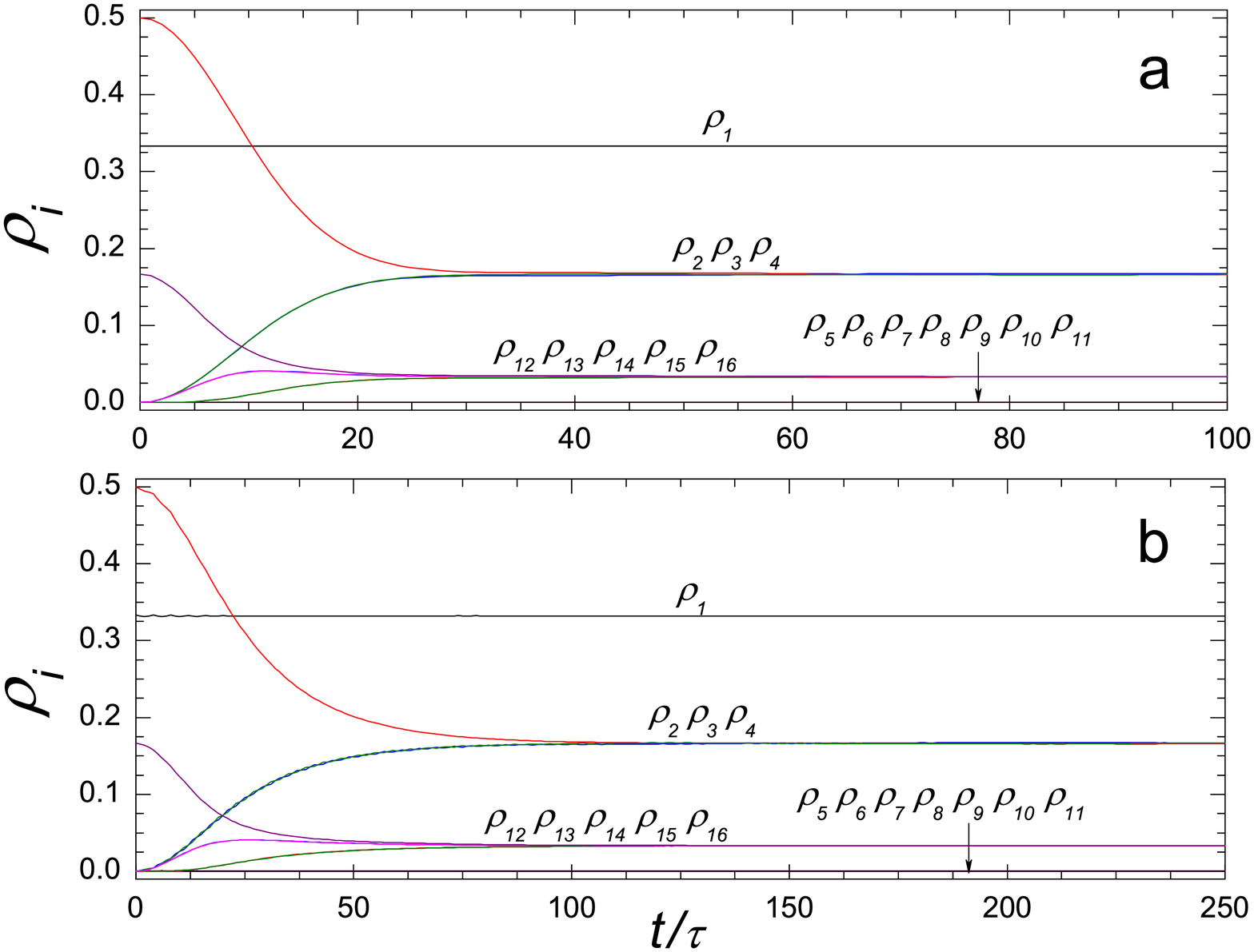}
\includegraphics[width=8.25cm]{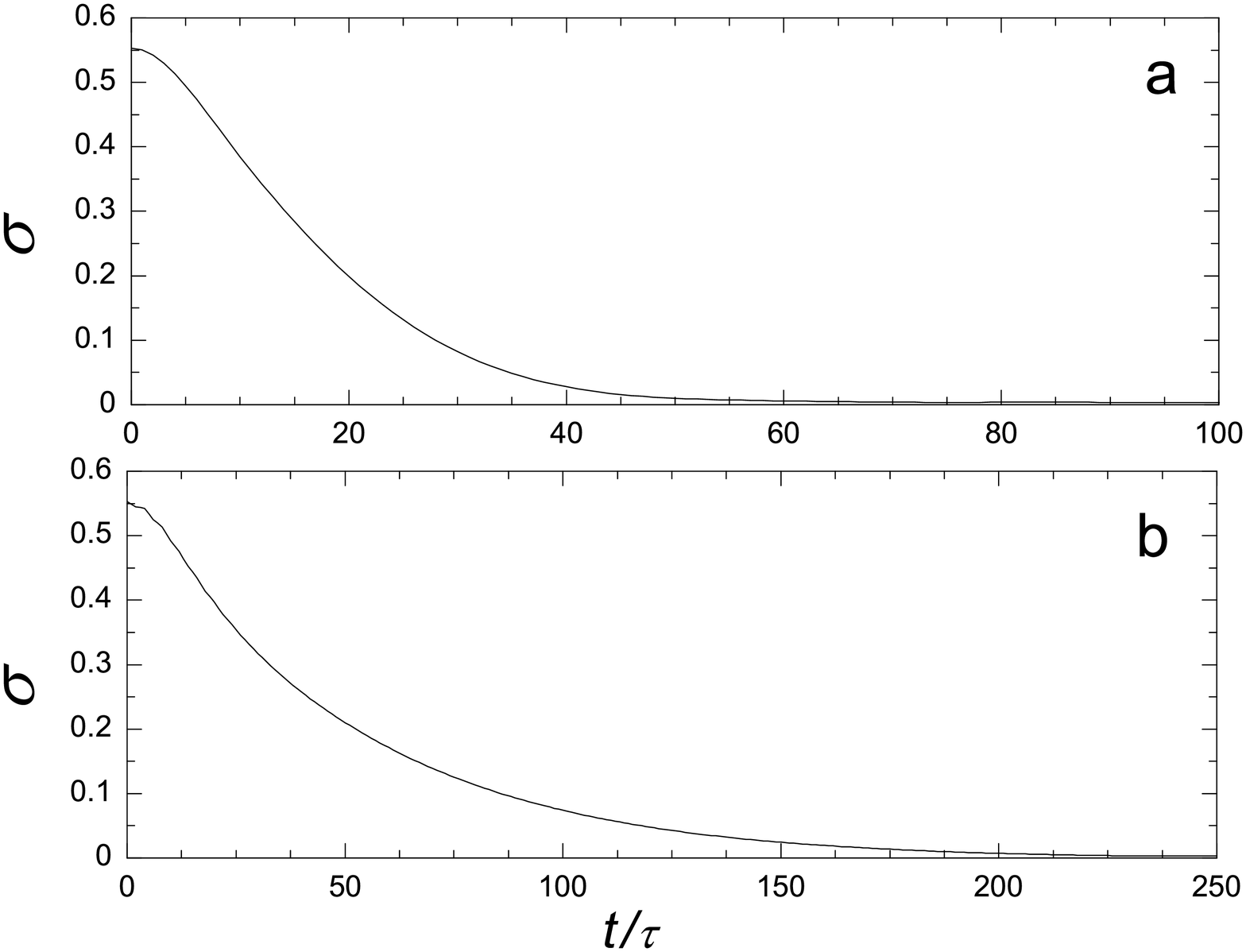}
}
\end{center}
\caption{Time evolution of the diagonal terms ($\protect\rho _{i}$) and sum
of the absolute values of the off-diagonal terms$\ (\protect\sigma )$ in the
reduced density matrix of a Heisenberg-ring $H_{S}$ ($J=-5$, $n_{S}=4$,
initial state $|UD\rangle _{S}$) coupled to a spin glass environment $H_{E}$
($\Omega =0.15$, $n=16$, initial state $|RANDOM\rangle _{E}$) via (\textbf{a}%
) Heisenberg interaction $H_{SE}$ ($\Delta =0.075$) or (\textbf{b})
Heisenberg-type interaction $H_{SE}$ ($\Delta =0.15$). Full decoherence is
observed in both cases, and the system $S$ relaxes to a state with equal
weights within each energy subspace, that is, $\protect\gamma \rightarrow 0$%
, a microcanonical ensemble. }
\label{c5fourconserved}
\end{figure*}

\begin{figure}[t]
\begin{center}
\includegraphics[width=8.25cm]{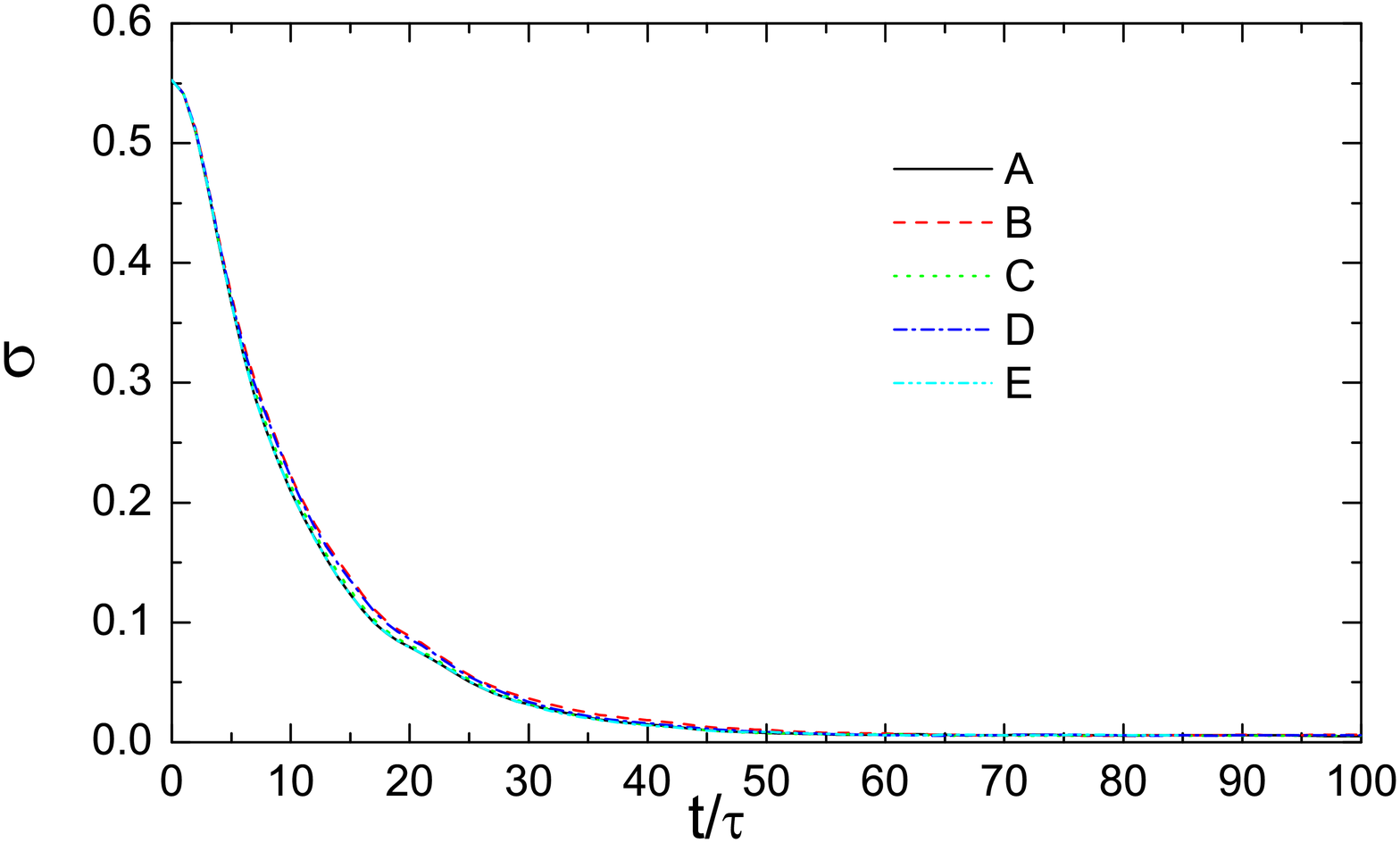}
\end{center}
\caption{The system represented here is the same as the one in Fig.~\protect
\ref{c5fourconserved}(a), except that the randomness in the coupling
constants of $H_{SE}$ or $H_{E}$, or the randomness in the initial state of
the environment, are different in the curves $A-E$. It is clear that the
time evolution of $\protect\sigma (t)$ is not sensitive to the different
random values of the coupling constants or the initial state of the
environment if they follow the same type of random distribution.}
\label{sigmadifferent}
\end{figure}

In the previous sections, we focused on the quantum system with only two
spins. Starting from this section we will consider more complicated systems
which contain more spins, e.g. four to eight spins.

In general when a quantum system interacts with a quantum environment, there
will be energy dissipation and entanglement of their wave functions. This
entanglement does not necessary lead to a classical mixed state of the
quantum system, especially if the environment has finite size. But
introducing certain properties, a finite quantum environment can also drive
the quantum system to an exact classical state. In the previous sections, we
showed that a frustrated environment can enhance the decoherence of a
two-spin system, and as we will see in what follows, this is also the case
for the many-spin system.

\subsection{Origin of the Microcanonical Distribution}

Let's first consider the case that there is no energy dissipation or the
energy dissipation is very small. In this case the decoherence of the
quantum system fully originates from the phase correlation of the
environment. In Fig.~\ref{c5fourconserved}, a Heisenberg-ring with four
spins is coupled to a frustrated spin glass consisting of 14 spins, and the
interaction between the quantum system and the environment is isotropic or
anisotropic but very small (comparing to the coupling within the system,
i.e. $\Delta =0.15\ll |J|=5$). In the former case the energy of the quantum
system is conserved, and in the latter case the energy dissipation is very
small so that it can be ignored. The system has four distinct eigenvalues ($%
E_{1}=-2$, $E_{2-4}=-1$, $E_{5-11}=0$, and $E_{12-16}=1$) and sixteen
different eigenstates. During the time-integration of the TDSE, the reduced
density matrix of the system is calculated every $\tau =\pi /10$. In both
cases, the diagonal terms of the reduced density matrix of the
Heisenberg-ring approach a stable value after an initial decay or increase,
and the off-diagonal terms are all zero ($\sigma \rightarrow 0$), which
means that in both cases, the four spin quantum system approaches a fully
decoherent state.

Another significant result is that the degenerated energy eigenstates have
the same weigth distribution in the fully decoherent state, i.e. $\gamma
\rightarrow 0$, indicating that the final system is a microcanonical state
in each eigenenergy subspace. The diagonal elements of the reduced density
matrix corresponding to degenerate eigenstates with zero weight in the
initial state are zero and remain so during the time revolution, see $\rho
_{5-11}$ in Fig.~\ref{c5fourconserved}. This is simply due to the
conservation of energy of the system, and due to the fact that the time
evolution operator $e^{-iHt}$ prevents changing the coefficients
corresponding to these eigenstates.

\begin{figure*}[t]
\begin{center}
\includegraphics[width=14cm]{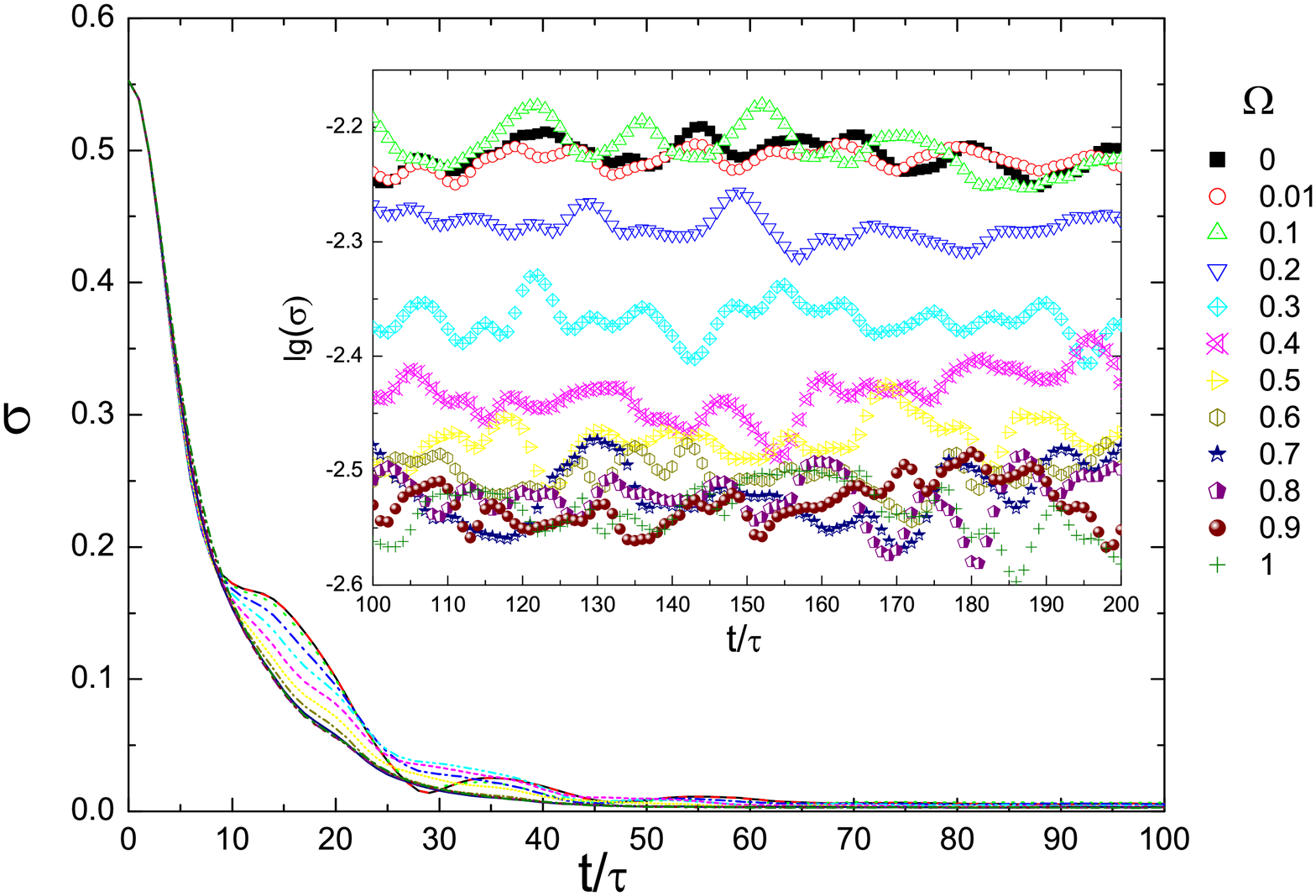}
\end{center}
\caption{Same as Fig.~\protect\ref{c5fourconserved}(a) except that the range
of the coupling strength ($\Omega $) in the environment is different.}
\label{sigmaomega}
\end{figure*}

\begin{figure}[t]
\begin{center}
\includegraphics[width=8.25cm]{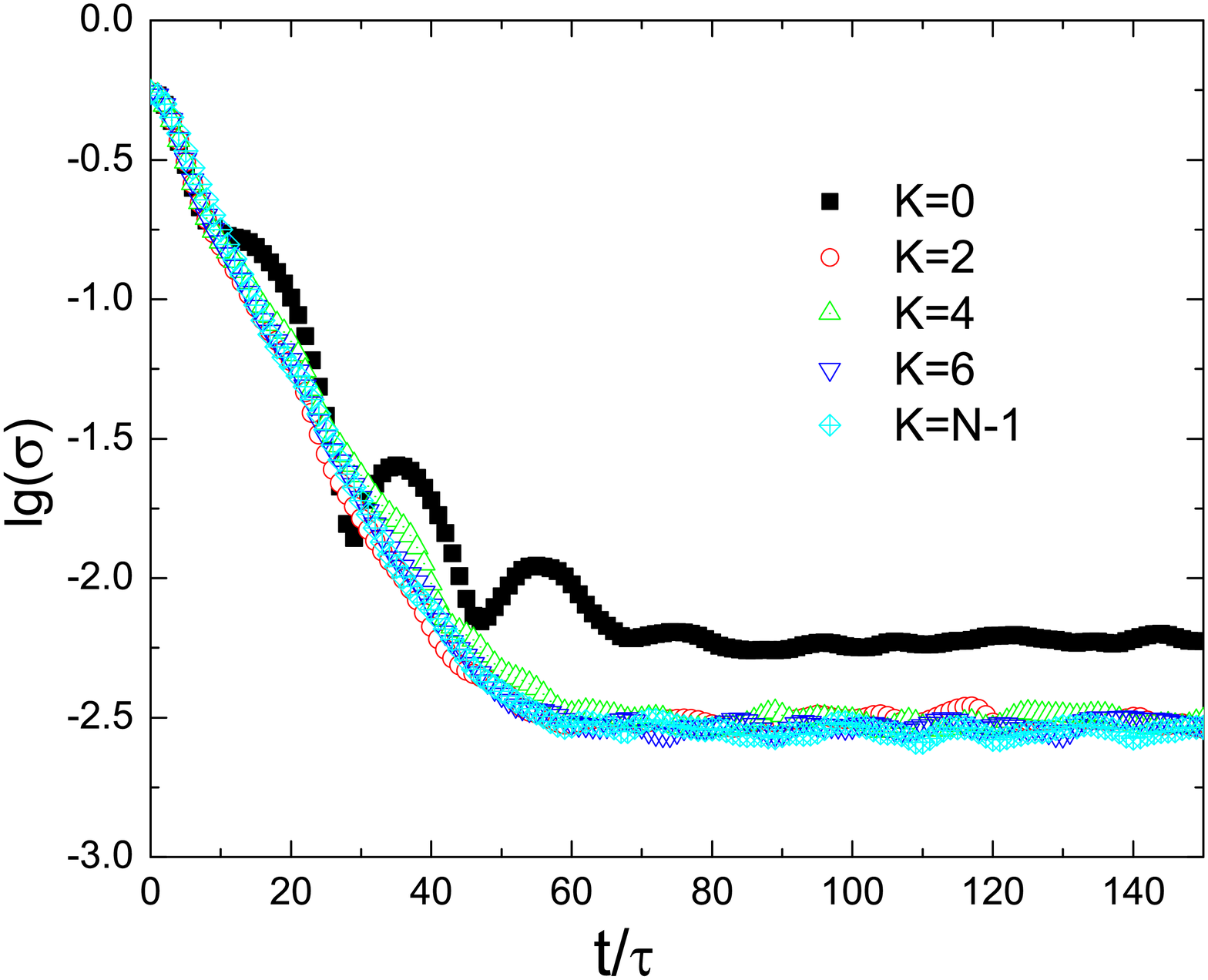}
\end{center}
\caption{Same as Fig.~\protect\ref{c5fourconserved}(a) except that the
topological structure (connectivity $K$) of the environment is different.}
\label{lgsigmadiffk}
\end{figure}

In Fig.~\ref{sigmadifferent}, we plot the time evolution of $\sigma \left(
t\right) $\ with different random realizations for the initial state of the
environment or for the model parameters $\Omega _{i,j}^{\alpha }$ and $%
\Delta _{i,j}^{\alpha }$. The difference between the curves is very small,
indicating that in our model, a particular randomness of the coupling
parameters or initial state is not relevant to the general properties of the
simulation results.

In Fig.~\ref{sigmaomega}, we show the time evolution of $\sigma \left(
t\right) $ for different coupling strengths ($\Omega $) in the environment.
In general, increasing the coupling strength within the environment will
increase the effective energy range of the bath, which leads the decoherence
more completely. But as we have shown in the case of the two-spin system,
the coupling strength should not be too large, otherwise the energy
resolution of the bath will be too small to lead the full decoherence of the
system.

In Fig.~\ref{lgsigmadiffk}, we show the time evolution of $\lg [\sigma
\left( t\right) ]$ for the same systems but with different topological
structures (connectivity $K$) in the environment. It is clear that as soon
as there is frustrated interaction within the environment ($K>0$), no matter
what kind of topological structure it is, the decoherence of the quantum
system is quite similar.

\subsection{Summary}

If there is no energy dissipation, or the energy dissipation is so small
that it can be ignored, then the entanglement between the quantum system and
the environment occurs only in the subspace of the (degenerate) eigenstates
which have nonzero weigth distribution in the initial state of the quantum
system. That is, the possible pointer states in the mixed states are
determined by the initial state of the quantum system itself. If a
particular environment can lead to the decoherence of the quantum system
without energy dissipation, then turning on the energy dissipation will
still lead to decoherence, and even more completely. In fact, energy
dissipation is not related to the question whether a quantum system can
evolve into a classical mixed state or not. The main difference between
decoherence with or without energy dissipation are the number of possible
pointer states in the mixed state. That is, an additional integral of motion
of the system will limit the number of pointer states, and therefore a full
decoherence state is a mixture with microcanonical distribution in each
eigenenergy subspace under the extra conservation law. On the contrary, if
there is enough energy dissipation between the two systems, then all
eigenstates of the quantum system are possible pointer states. And more
importantly, as we will show in the next section, the mixed state of the
quantum system follows the canonical distribution \cite{Yuan2009}.

\section{Thermalization of a Many-Spin System}

In the previous section, we have shown that turning on the interaction
between the many-spin system and the environment, leads to a reduction of
the coherence in the quantum system. The coupling with the environment
causes the initial pure state of the quantum system to evolve into a mixed
state, obtained by tracing out all the degrees of freedom of the
environment. The pointer states in the mixed state are determined by the
initial state of the quantum system if there is an additional integral of
motion, e.g.. the conserved energy or magnetization. This leads to a
microcanonical ensemble under a certain conservation law. On the other hand,
if there is enough energy dissipation without any additional integral of
motion, we expect that the mixed state is a canonical ensemble \cite%
{Yuan2009}.

Earlier demonstrations of the fact that the system can be in the canonical
ensemble state are based on Ergodic averages \cite%
{Bocchieri1959,Shankar1985,Tasaki1998,sait96} or canonical typicality~\cite%
{Popescu2006,Rigol2008,Goldstein2006,Reimann2007,Reimann2008,Gemmer2006,Gemmer2006b}%
. The Ergodic averages consider the dynamics of a closed quantum system, and
prove that in certain quantum systems, the expectation values of the
dynamical variables of the system approach their values for the subsystem
that is in the thermal equilibrium state. This is similar to the assumption
of classical statistical physics, that is, during large enough time, the
trajectory of the many particle system in the phase space will pass all
possible points, and therefore the average of these points in the phase
space will follow a certain distribution. On the contrary, the canonical
typicality does not consider the dynamics of the system but assumes that if
the whole system is in the microcanonical ensemble then its subsystems are
in the canonical ensemble. The statement of the canonical typicality is
quite general, but it is clear that the dynamical procedure is missing.
Moreover, the coupling between the quantum system and the environment is
assumed to be so small that it can be neglected in the theory. Therefore it
is important to show how the canonical distribution of statistical mechanics
relates to the dynamical evolution of the quantum system.

\subsection{Origin of the Canonical Ensemble}

\begin{figure*}[t]
\begin{center}
\includegraphics[width=14cm]{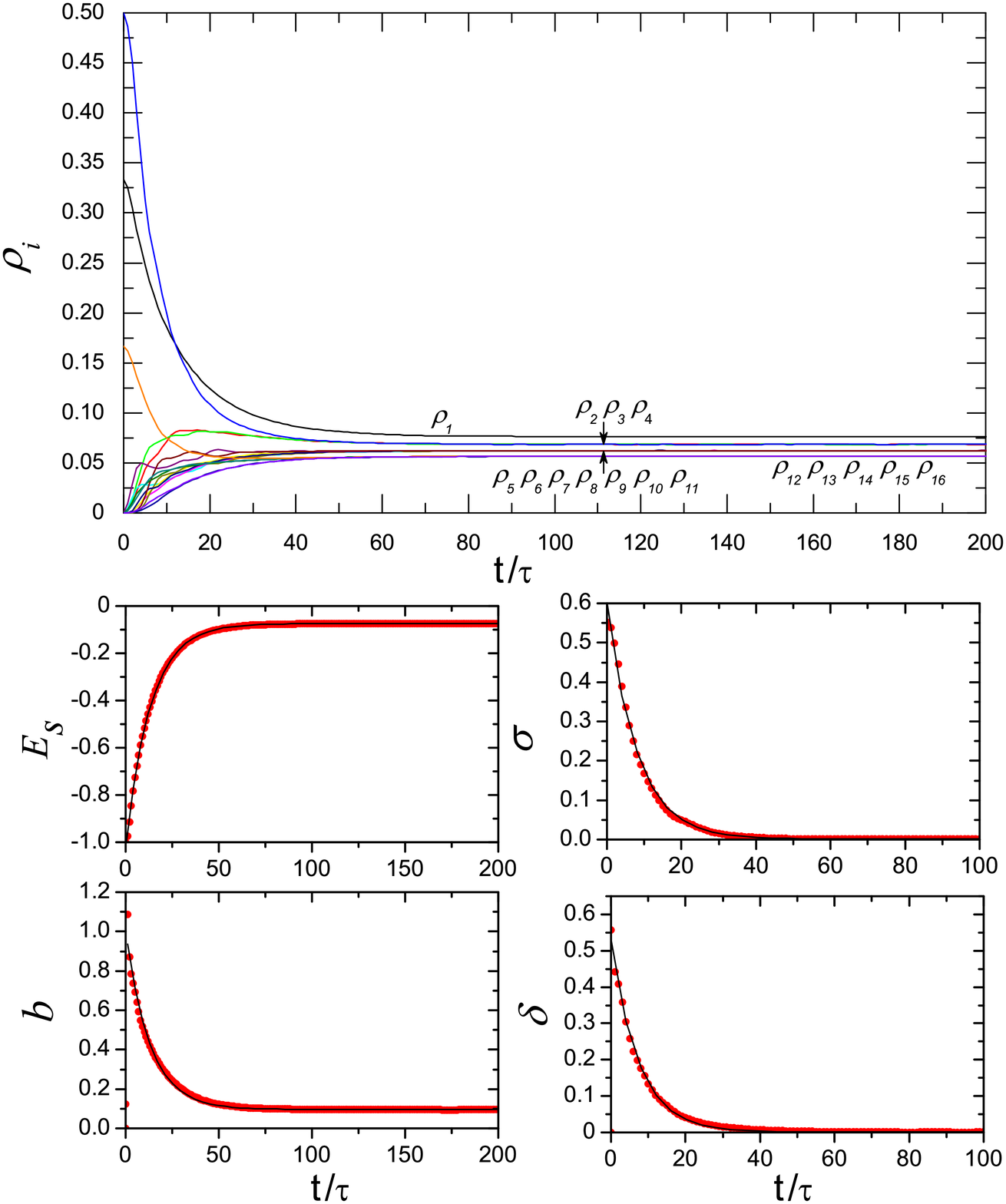}
\end{center}
\caption{Simulation results for the diagonal elements $\protect\rho %
_{i}\equiv \widehat{\protect\rho }_{ii}(t)$ of the density matrix of $S$,
the energy $E_{S}\equiv E_{S}(t)$, the effective inverse temperature $%
b\equiv b(t)$ and its variance $\protect\delta \equiv \protect\delta (t)$,
and $\protect\sigma \equiv \protect\sigma (t)$ which is a measure for the
decoherence in $S$, as obtained by solving the TDSE for the whole system
with a Heisenberg-ring $H_{S}$ ($J=-1$, $n_{S}=4$), a Heisenberg-type
interaction $H_{SE}$ ($\Delta =0.3$), a spin glass environment $H_{E}$ ($%
\Omega =1$, $n=18$), and $\protect\tau =\protect\pi /10$. The initial state
of the whole system is a product state $\left\vert UD\right\rangle
_{S}\otimes \left\vert RANDOM\right\rangle _{E}$. The red dots in the small
panels represent the simulation data, and the black curves are fitting
curves (see text).}
\label{canonicalfig1}
\end{figure*}

As a frustrated environment is very effective for creating full decoherence (%
$\sigma \rightarrow 0$) in a quantum spin system, and full decoherence is a
necessary condition for the state of the system to converge to its canonical
distribution, we have chosen spin glass environments, which have no obvious
symmetries \cite{Yuan2009}.

\begin{figure}[t]
\begin{center}
\includegraphics[width=8.25cm]{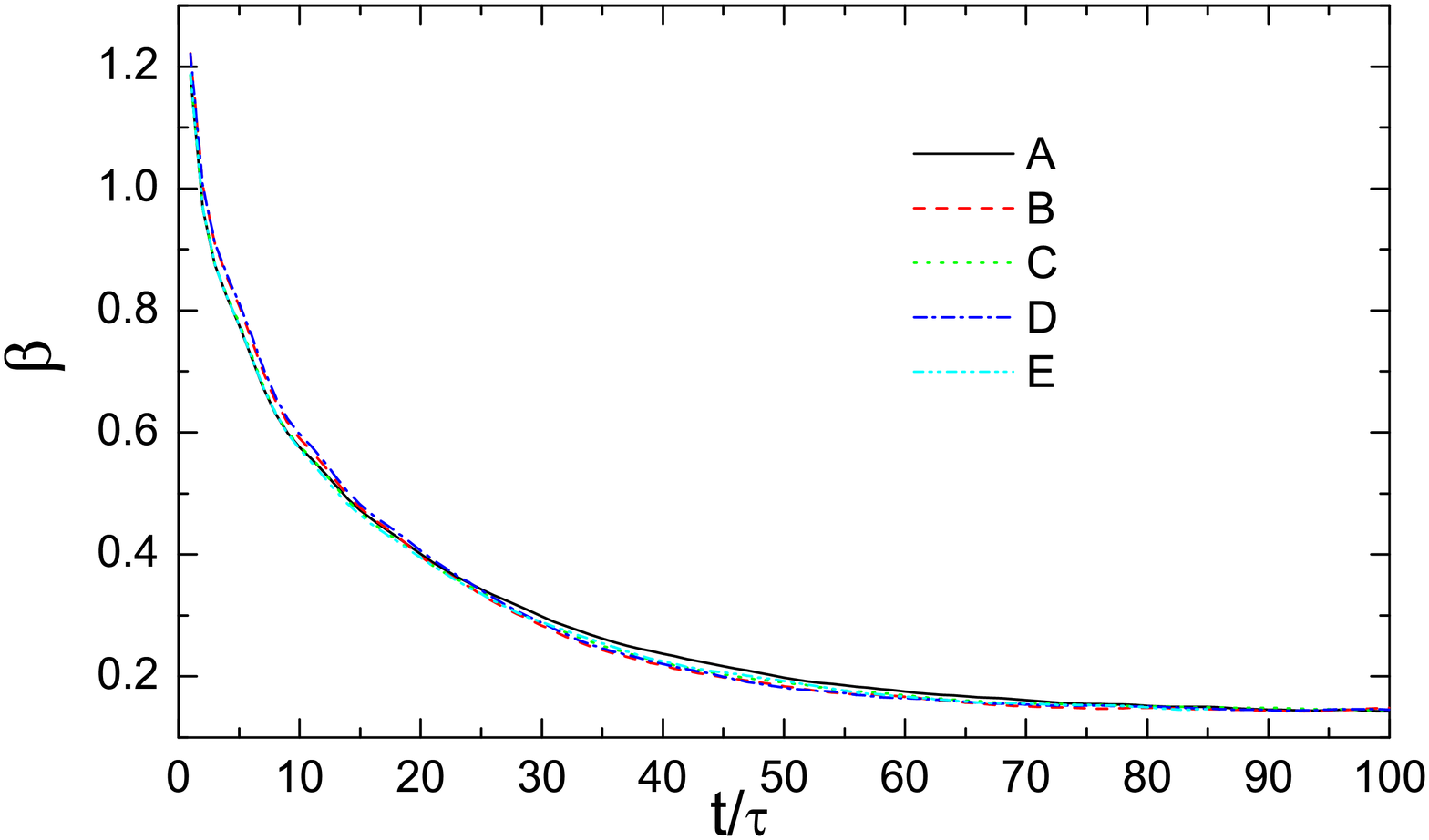}
\end{center}
\caption{Same as Fig.~\protect\ref{canonicalfig1}, except that the
randomness in the coupling constants of $H_{SE}$ or $H_{E}$, or the
randomness in the initial state of the environment, are different in the
curves $A-E$. It is clear that the time evolution of the effective
temperature $b(t)$ is not sensitive to the different random values of the
coupling constants or the initial state of the environment if they follow
the same type of random distribution.}
\label{betadifferent}
\end{figure}

\begin{figure}[t]
\begin{center}
\includegraphics[width=8.25cm]{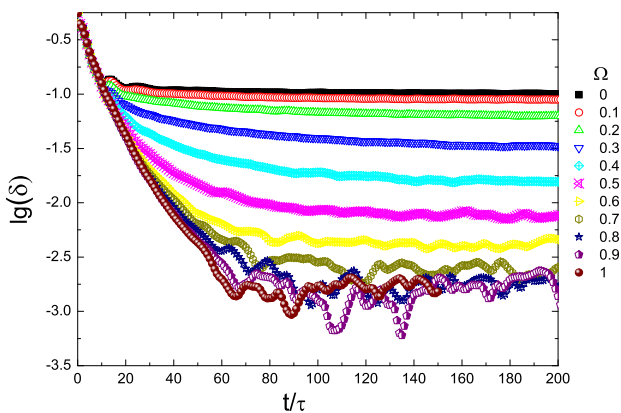}
\end{center}
\caption{Same as Fig.~\protect\ref{canonicalfig1} except that the range of
the coupling strength ($\Omega $) is different in the environment of each
curve.}
\label{lgdeltaomega}
\end{figure}

\begin{figure}[t]
\begin{center}
\includegraphics[width=8.25cm]{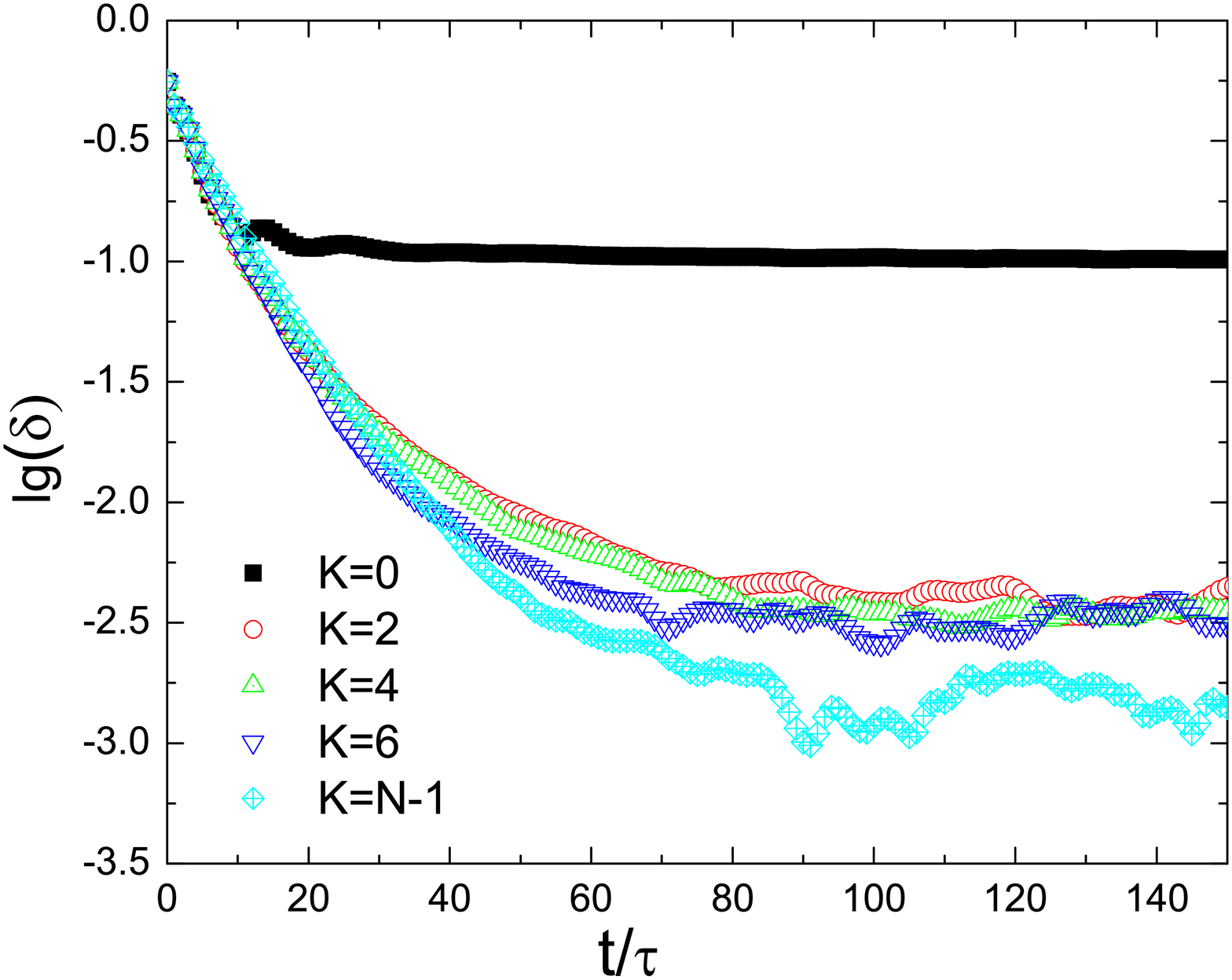}
\end{center}
\caption{Same as Fig.~\protect\ref{canonicalfig1} except that the
topological structure (connectivity $K$) of the environment is different. }
\label{lgdeltadiffk}
\end{figure}

First, we consider a system (Heisenberg-ring $H_{S}$) interacting
(Heisenberg-type $H_{SE}$) with an environment (spin glass $H_{E}$). The
system has four distinct eigenvalues ($E_{1}=-2$, $E_{2-4}=-1$, $E_{5-11}=0$%
, and $E_{12-16}=1$) and sixteen different eigenstates. The environment has $%
2^{18}$ eigenstates. During the time-integration of the TDSE, the reduced
density matrix of the system is calculated every $\tau =\pi /10$ as in the
previous section As we described earlier, the values of the diagonal
elements $\widehat{\rho }_{ii}$ yield an estimate for the effective inverse
temperature $b(t)$, the error $\delta (t)$ for this estimate and the measure 
$\sigma (t)$ for the deviation from a non-diagonal matrix. The energy of the
system is obtained as $E_{S}(t)=\mathbf{Tr}_{S}\widehat{\rho }(t)H_{S}$.

\begin{figure}[t]
\begin{center}
\includegraphics[width=8.25cm]{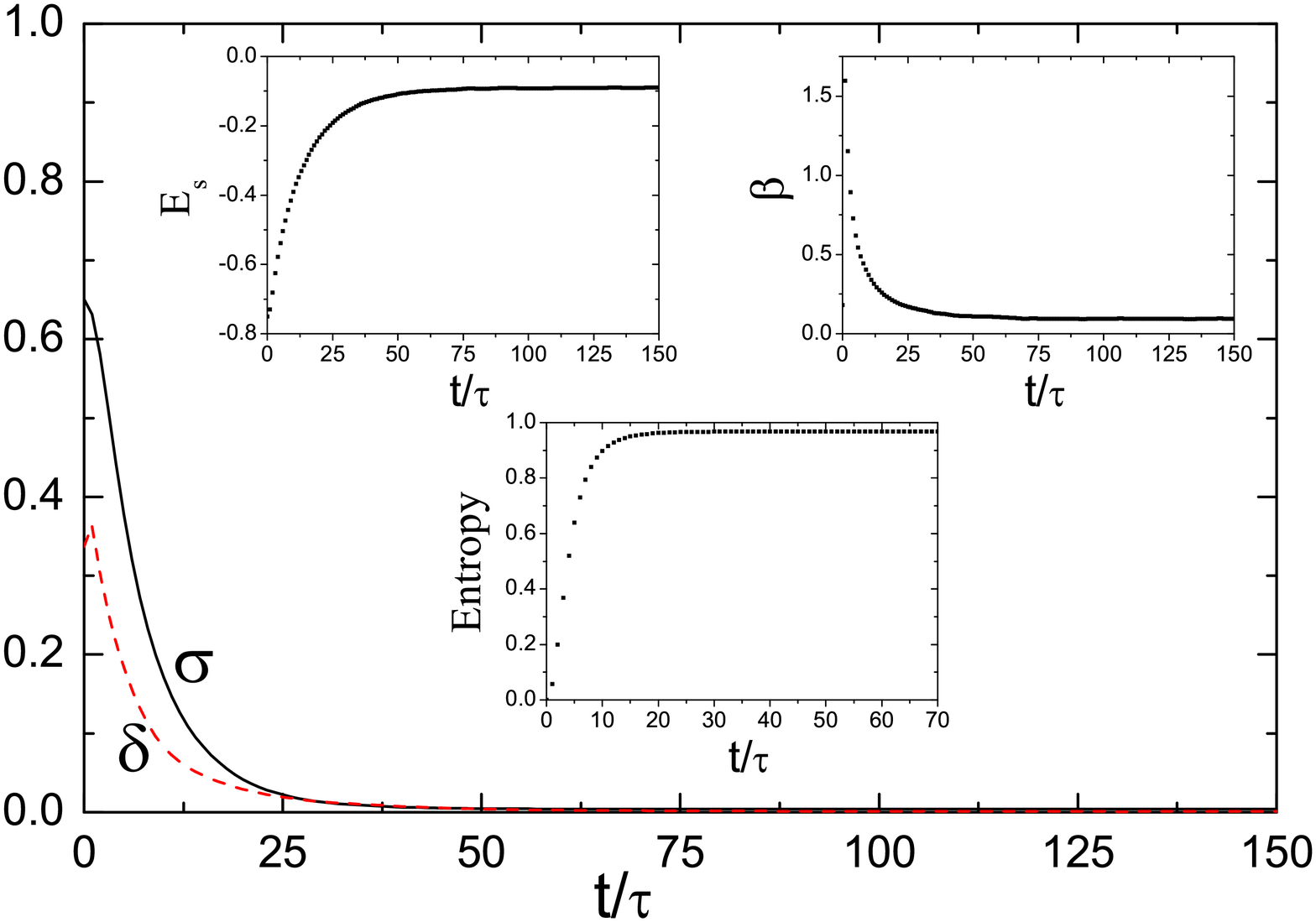}
\end{center}
\caption{Same as Fig.~\protect\ref{canonicalfig1} except that $n_{S}=5$ and $%
n=17$.}
\label{canonicalfig1a}
\end{figure}

\begin{figure*}[t]
\begin{center}
\includegraphics[width=16cm]{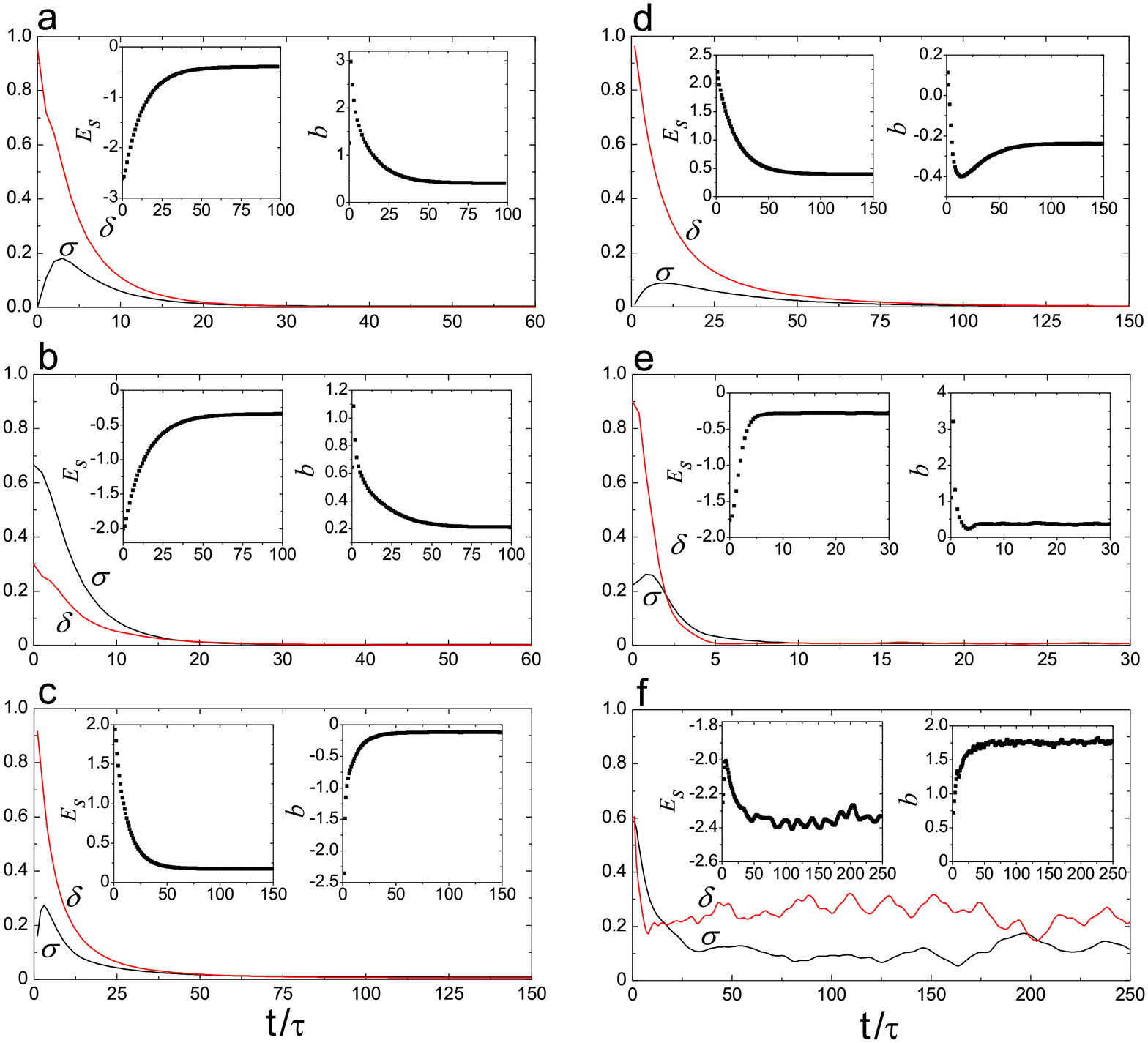}
\end{center}
\caption{ Simulation results for the energy $E_{S}\equiv E_{S}(t)$, the
effective inverse temperature $b\equiv b(t)$, its variance $\protect\delta %
\equiv \protect\delta (t)$, and the deviation from a diagonal matrix $%
\protect\sigma \equiv \protect\sigma (t)$ as obtained by the solution of the
TDSE for a variety of different systems $S$ coupled to a spin glass
environment $H_{E}$ via a Heisenberg-type interaction $H_{SE}$. The systems
used are \textbf{a}: XY-ring, \textbf{b} and \textbf{f}: Heisenberg-ring, 
\textbf{c}: Ising-ring, \textbf{d}: Heisenberg-triangular-lattice, and 
\textbf{e}: spin glass. The initial states of the whole system are \textbf{a}%
: $|GROUND\rangle _{S}\otimes |RANDOM\rangle _{E}$, \textbf{b}: $|UD\rangle
_{S}\otimes |RANDOM\rangle _{E}$, \textbf{c}: $|UU\rangle _{S}\otimes
|RR\rangle _{E}$, \textbf{d}: $|UU\rangle _{S}\otimes |RANDOM\rangle _{E}$, 
\textbf{e}: $\left\vert GROUND\right\rangle _{S}\otimes |UD\rangle _{E}$,
and \textbf{f}: $\widetilde{|UD\rangle }_{S}\otimes |GROUND\rangle _{E}$.
The numbers of spins in the system are $n_{S}=8$ for cases \textbf{a}-%
\textbf{c} and $n_{S}=6$ for cases \textbf{d}-\textbf{f}. The number of
spins in the environment is $n=16$ for all cases. The model parameters are $%
J=-1$, $\Delta =0.3$ and $\Omega =1$, except for case \textbf{e} in which $%
\Delta =1$.}
\label{canonicalfig2}
\end{figure*}

From the simulation results, shown in Fig.~\ref{canonicalfig1}, it is clear
that for $t>50\tau $, each diagonal element $\widehat{\rho }_{ii}$ of the
reduced density matrix converges to one out of four stationary values,
corresponding to the four non-degenerate energy levels of the system. This
convergence is a two-step process. First the system looses all coherence, as
indicated by the vanishing of $\sigma \left( t\right) $ for $t>50\tau $. The
time dependence of $\sigma \left( t\right) $ fits very well to an
exponential law 
\begin{equation}
\sigma \left( t\right) =\sigma _{\infty }+Ae^{-t/T_{2}},
\end{equation}%
with $\sigma _{\infty }=0.00128$, $A=0.602$ and $T_{2}=8.01\tau $. In the
small panels of Fig.~\ref{canonicalfig1}, the red dots are the simulation
data and the black curves are the fitting function. Likewise, the vanishing
of $\delta (t)$ on the same time-scale ($T_{2}=7.32\tau $), indicates that
the density matrix of the system converges to the canonical distribution
with the same speed of decoherence.

The effective temperature $b(t)$ and the energy of the system $E_{S}\left(
t\right) $ also fit very well to the exponential laws 
\begin{equation}
b\left( t\right) =\beta +Be^{-t/T_{1}},
\end{equation}%
and 
\begin{equation}
E\left( t\right) =E_{\infty }+Ce^{-t/T_{1}},
\end{equation}%
with $\beta =0.0962$, $B=-0.900$, and $T_{1}=13.3\tau $ and $E_{\infty
}=-0.0745$, $C=-0.952$. The estimated values for $T_{1}$ and $T_{2}$ change
very little if we choose different random realizations for the initial state
of the environment or for the model parameters $\Omega _{i,j}^{\alpha }$ and 
$\Delta _{i,j}^{\alpha }$ (see Fig.~\ref{betadifferent}). If we change their
range, $T_{1}$ and $T_{2}$ also change, as naively expected.

In order to verify the role of the dynamics within the bath to the
thermalization of the quantum system, we plot the time evolution of $\delta
\left( t\right) $ for different coupling strength ($\Omega $) in Fig.~\ref%
{lgdeltaomega}. Similar as the dependence of $\sigma $\ on $\Omega $ in Fig.~%
\ref{sigmaomega}, increasing the coupling strength (within a certain range)
will increase the effective energy range of the bath, which leads to a more
complete decoherence and thermalization.

\begin{figure}[t]
\begin{center}
\includegraphics[width=8.25cm]{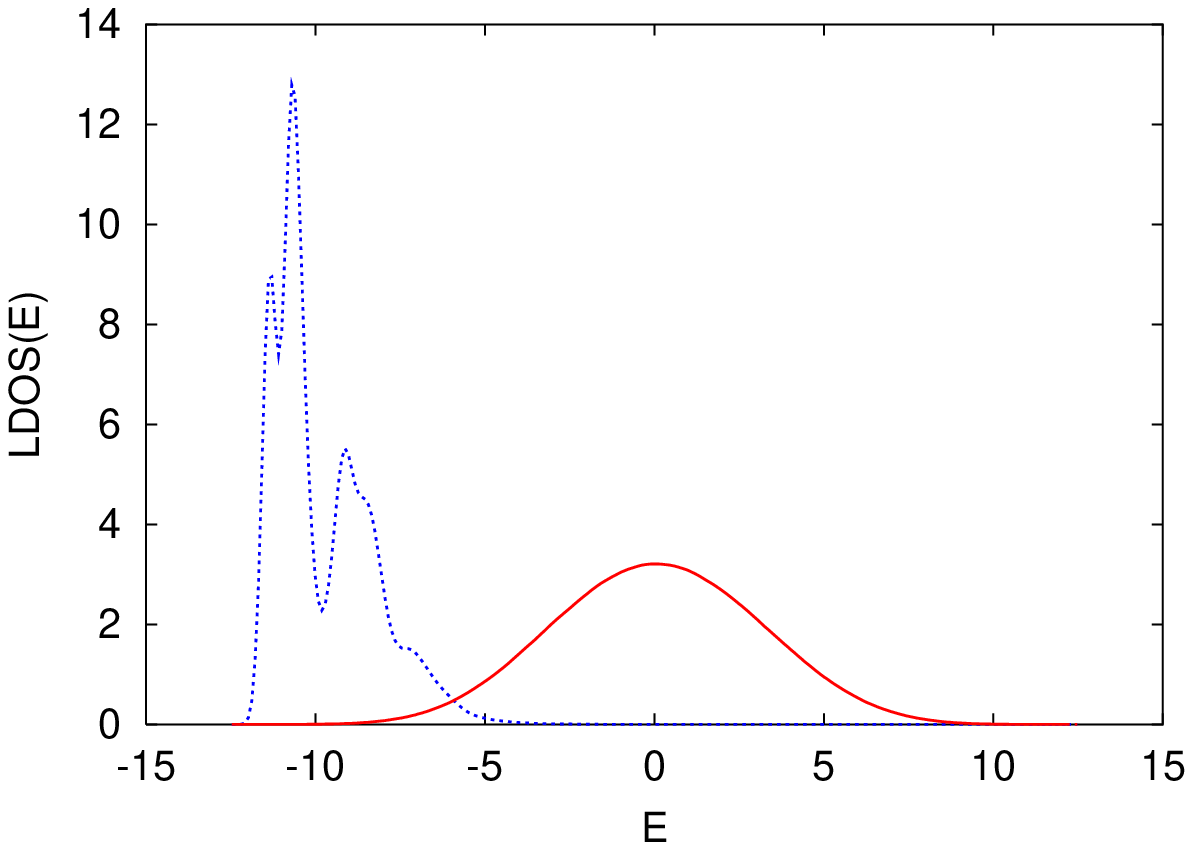}
\end{center}
\caption{Simulation results for the local density of states as a function of
the energy. Solid line: Case corresponding to Fig.~\protect\ref%
{canonicalfig2}\textbf{b}. The initial state is $|UD\rangle _{S}\otimes
|RANDOM\rangle _{E}$; Dashed line: Case corresponding to Fig.~\protect\ref%
{canonicalfig2}\textbf{f}. The initial state is $|UD\rangle _{S}\otimes
|GROUND\rangle _{E}$. Adapted from Ref. \protect\cite{Yuan2009}.}
\label{canonicalfig3}
\end{figure}

The simulation results of a similar system with one extra spin in the
quantum system ($n_{s}=5$) and one spin less in the environment $(n=17$) are
shown in Fig.~\ref{canonicalfig1a}, and are very similar to the ones shown
in Fig.~\ref{canonicalfig1}. These simulations demonstrate that the system
first looses all coherence and then, on a longer time-scale, relaxes to its
thermal equilibrium state with a finite temperature. In terms of the theory
of magnetic resonance~\cite{Abragam61}, $T_{1}$ and $T_{2}$ are the times of
dissipation and dephasing, respectively. In the case of very small $H_{E}$,
one should expect, instead of an exponential decay of $\sigma $ and $E$, a
Gaussian decay, as observed in Ref.~\cite{Yuan2006,Yuan2007,Yuan2008}.

It is necessary to extend the types of interaction (Hamiltonian) to verify
the generality of the above results. The time evolution of $\delta \left(
t\right) $ for the same systems but with different topological structures
(connectivity $K$) in the environment are shown in Fig.~\ref{lgdeltadiffk}.
Increasing the connectivity in the environment will increase the energy
resolution and make the dynamics in the environment more complicated, which
enhances the approach to the canonical distribution of the system. This is
quite different from the effect of the topological structure on the
decoherence shown in Fig.~\ref{lgsigmadiffk}, where as soon as there is
frustrated interaction within the system ($K>0$), the decoherence of the
quantum system is quite similar for different topological structures.

More results for the system with different symmetries and connectivities but
with the same type of environments ($H_{E}$) and the same type of
interactions ($H_{SE}$) are shown in Fig.~\ref{canonicalfig2}. The systems
used are a XY-ring, a Heisenberg-ring, an Ising-ring, a
Heisenberg-triangular-lattice, and a spin glass. From the results
represented in Fig.~\ref{canonicalfig2}, it is clear that independent of the
internal symmetries and the connectivity of the system, and independent of
the initial state of the whole system, all systems relax to a state with
full decoherence, except case \textbf{f}. The main difference between case 
\textbf{f }and the other cases is the initial state of the environment.
Since the environment in our model is a highly frustrated system, the LDOS
of the whole system covers the whole spectrum of the energy space, no matter
the initial state of the environmental spins is all spins up or all spins
down, or random spins up and down, or a random superposition of all the
states in the spin up-down basis. But if the environment is prepared in the
ground state or near ground state, then the LDOS of the whole system becomes
more sharply peaked, see \textbf{b} and \textbf{f} in Fig.~\ref%
{canonicalfig3}. Up to a trivial normalization factor, the LDOS curve for
case \textbf{b} is indistinguishable from the density of states (data not
shown) calculated from the solution of the TDSE using the technique
described in Ref.~\cite{HAMS00}. This suggests that if the environment
starts from the random superposition of all its energy eigenstates, all
states of the whole system may participate in the decoherence/relaxation
process. In contrast, the LDOS curve for case \textbf{f} has a very small
overlap with the density of states. Therefore, starting with an environment
in the ground state, only a relatively small number of states participates
in the decoherence process, as confirmed by the results for $\sigma (t)$
shown in Fig.~{\ref{canonicalfig2}}\textbf{f}.

One should also notice that in case \textbf{b}, $\sigma $ vanishes
exponentially with time, whereas in the other cases (\textbf{a},\textbf{c},%
\textbf{d},\textbf{e}), $\sigma $ initially increases and then vanishes
exponentially with time, due to the entanglement between the system and the
environment. This observation is in concert with our earlier work~\cite%
{Yuan2006,Yuan2007,Yuan2008}. Furthermore, in all cases except \textbf{f},
the system always relaxes to a canonical distribution ($\delta \rightarrow 0$%
) as soon as it is in the state with full decoherence ($\sigma \rightarrow 0$%
), indicating that the time of decoherence ($T_2$) and the time of
thermalization ($T_1$) is almost the same. In agreement with the results
depicted in Fig.~\ref{canonicalfig1}, the decoherence time $T_{2}$ is
shorter than the typical time scale $T_{1}$ on which the system and the
environment exchange energy. Note that in contrast to the cases considered
in the theory of nuclear magnetic resonance, in most of our simulations, $%
H_{S}$, $H_{E}$ and $H_{SE}$ are comparable so the standard perturbation
derivation of $\sigma $ and $E$ does not work.

The negative temperature ($b<0$) is also observed in \textbf{c} and \textbf{d%
}$.$ In fact, as the temperature $T$ is defined as 
\begin{equation}
\frac{1}{T}=\frac{dS}{dE},
\end{equation}%
where $S$ and $E$ are the entropy and energy. In a quantum spin system, the
entropy may decrease when the energy increases. For example, the states with
all spins up and all spins down have the same entropy, but can have totally
different energy. Suppose there is no interaction between $N $ ferromagnetic
spins ($J>0$), and there is a uniform magnetic filed applied on the $Z+$
direction, then the state with all spins up is the ground state, and with
all spins down is the eigenstate with highest energy. Changing the
magnetization of the states from $M_{\max }=N/2$ to $M_{\min }=-N/2$ will
change the sign of the temperature $T(M)$ at the point $M=0$, that is, $T>0$
when $M>0$ and $T<0$ when $M<0$.

\subsection{Summary}

We have shown that if we have a system that interacts with an environment
and the whole system forms a closed quantum system that evolves in time
according to the TDSE, then a frustrated environment with a random
distribution in the energy basis will lead to the full decoherence of the
system. Furthermore, if the system and environment can exchange energy, the
range of energies of the environment is large compared to the range of
energies of the system, then the mixed state of the system is a canonical
distribution.

\section{Conclusion and Discussion}

The results presented here have been obtained from an \emph{ab initio}
numerical solution of the TDSE in the absence of, for instance, dissipative
mechanisms, and demonstrate that the existence of the microcanonical
distribution (in each eigenenergy subspace) and the canonical ensemble is a
direct consequence of quantum dynamics.

We emphasize that our conclusion does not rely on time averaging of
observables, in concert with the fact that real measurements of
thermodynamic properties yield instantaneous, not time-averaged, values.
Furthermore and perhaps a little counter intuitive, our results show that
relatively small environments ($\approx 20$ spins) are sufficient to drive
the system ($2-8$ spins) to thermal equilibrium and that there is no need to
assume that the interaction between the system and environment is weak, as
is usually done in kinetic theory. Note that even if most cases shown in
this review clearly indicate a full decoherence of the quantum system, it
does not mean that the condition of full decoherence is a nontrivial
requirement.

To conclude, we find that:

(1) Frustration of the interactions, either within the environment or
between the system and the environment, enhances the decoherence of the
quantum system.

(2) The quantum system more easily evolves into its ground state when the
latter is more degenerate or less entangled, or has certain additional
integrals of motion.

(3) The distribution of the state of a quantum system is the microcanonical
or canonical ensemble only if the system is in a fully decoherent state.

(4) The restriction of a fully decoherent state to be a microcanonical
ensemble per eigenenergy subspace is the presence of an additional integral
of motion except a conserved energy. For example, a conserved magnetization
of the quantum system prevents parts of the degenerate eigenstates to be the
pointer states.

(5) The restriction of a microcanonical ensemble to be a canonical ensemble
is the presence of an additional integral of motion, so that the energy of
each subspace is conserved.

(6) The distributions in quantum statistical mechanics, such as the
microcanonical and canonical distributions, are the direct consequence of
quantum dynamics.

Finally we want to discuss the second law of thermodynamics in quantum
systems.

The second law of thermodynamics states that the entropy of an isolated
system which is not in equilibrium tends to increase over time, approaching
a maximum value at equilibrium. In quantum mechanics, the state of a closed
quantum system is always a pure state and therefore its entropy is a
constant (zero). It is thus clear that the second law of thermodynamics is
not valid in a closed quantum system. If a quantum system starts to interact
with an environment, its entropy increases from zero but may not reach a
maximum value at equilibrium. The dynamics of the whole system could be
periodical and therefore the time evolution of the states could be
reversible.

Our numerical results show that if the quantum system becomes a classical
mixed state, then the time evolution becomes irreversible and the entropy
becomes stable when it reaches the maximum value. In fact, the state with
the microcanonical distribution in each eigenenergy subspace or in the
canonical ensemble has the maximum entropy within all possible states that
the system could be. For a quantum system with a certain energy and a fixed
number of particles it is such that, if there is a conservation law to
restrict some eigenstates in the reduced density matrix, then the state with
maximum entropy corresponds to the one with all the accessible degenerate
states having the same weigth distribution, i.e. a microcanonical
distribution per eigenenergy subspace. On the other hand, if there is no
such restriction on the eigenstates, then the canonical ensemble is the
state with maximum entropy, as proved by Jaynes in Ref.~\cite{Jaynes1957}.

We may conclude that the validity of the second law of thermodynamics in
quantum mechanics is related to the decoherence process of the quantum
system. If a quantum system becomes classical under the influence of the
environment, then its entropy will increase until it reaches a maximum value
of all possible mixed states, i.e. the microcanonical distribution per
eigenenergy subspace or the canonical ensemble. If a quantum system cannot
evolve into a stable mixed state, then its entropy will not always follow
the second law of thermodynamics.

\section{Acknowledgement}

It is a pleasure to thank H. De Raedt, M. I. Katsnelson, S. Miyashita, K.
Michielsen, F. Jin and S. Zhao for many helpful discussions. The support by
the Stichting Fundamenteel Onderzoek der Materie (FOM) and the Netherlands
National Computing Facilities foundation (NCF) are acknowledged.

\end{document}